\documentclass[11pt,draftcls,onecolumn]{IEEEtran}
\IEEEoverridecommandlockouts
\usepackage[doublespacing]{setspace}
\usepackage[cmex10]{amsmath}
\usepackage{textcomp}
\usepackage{amssymb}

\usepackage{nccmath}
\usepackage{cite}
\usepackage{graphicx}
\usepackage{color}

\usepackage{mathtools}

\newtheorem{theorem}{Theorem}
\newtheorem{lemma}{Lemma}

\begin{document}
\begin{singlespace}
\title{On Achievable Degrees of Freedom for the Frequency-Selective $K$-User Interference
Channel in the Presence of an Instantaneous Relay} 

\author{\normalsize Ali H. Abdollahi Bafghi,  Mahtab Mirmohseni, and Masoumeh Nasiri-Kenari\\

\thanks{
Ali H. Abdollahi Bafghi,  Mahtab Mirmohseni, and Masoumeh Nasiri-Kenari  are with the Department of Electrical Engineering, Sharif University of Technology, Tehran, Iran (email: aliabdolahi@ee.sharif.edu, mirmohseni@sharif.edu, mnasiri@sharif.edu). 

}

}

\maketitle

\begin{abstract}
In this paper, we study the degrees of freedom (DoF) of the frequency-selective $K$-user interference channel in the presence of an instantaneous relay (IR) with multiple receive and transmit antennas. We investigate two scenarios based on the IR antennas' cooperation ability. First, we assume that the IR receive and transmit antennas can coordinate with each other, where the transmitted signal of each transmit antenna can depend on the received signals of all receive antennas,
and we derive an achievable DoF for this model. In our interference alignment scheme, we divide receivers into two groups, called clean and dirty receivers. We design our scheme such that a part of the message of clean receivers can be demultiplexed at the IR. Thus, the IR can use these message streams for interference cancelation at the clean receivers. Next, we consider an IR, whose antennas do not have coordination with each other, where the transmitted signal of each transmit antenna only depends on the received signal of its corresponding receive antenna,
and we derive an achievable DoF for it. We show that the achievable DoF decreases considerably compared with the coordinated case. In both of these models, our schemes achieve the maximum $K$ DoFs, if the number of transmit and receive antennas is more than a finite threshold.
\end{abstract}


\section{Introduction}
Spectrum sharing in wireless networks seems to be an inevitable solution for the increasing bandwidth demands. How to treat interference is one of the main challenges in these scenarios. Interference alignment is proved to be a useful solution that aligns all the interference signals into a smaller subspace, allowing the remaining signal space to be used for the transmission of main signals. Thereby, it can achieve the maximum degrees of freedom (DoF) of $\frac{K}{2}$ in a $K$-user interference channel \cite{Cadambe1}. An interesting question would be to find tools that can improve this maximum value for the DoF. Instantaneous relay (IR) is one of these tools \cite{El Gamal,Lee}.

In an IR, the transmitted signal in the $i$-th time slot is a function of all received signals from the first time slot up to the current ($i$-th) time slot, while in the classic relay, the transmitted signal in the $i$-th time slot does not depend on the received signal in the $i$-th (current) time slot\footnote{It has been shown in \cite{Jafar6} that a classic relay cannot increase the DoF of the $K$-user interference channel.}.
Though for the current technology, IR might seem impractical, there are significant results on the IR, and also the active intelligent reflecting surface is a promising technology that makes it possible to realize an IR in the near future \cite{Renzo}.

The capacity of wireless networks in the presence of an IR has been studied in \cite{El Gamal2}-\cite{Azari}.
El Gamal, et al, in \cite{El Gamal2} showed that in the presence of an IR, rates higher than the existing cut-set bound for the classic relay can be achieved for a point-to-point channel.
In \cite{salimi} a new upper bound was derived for the capacity of the channel with an IR.
The authors in \cite{Chang} studied the two-user interference channel in the presence of an IR and derived an outer bound for the Gaussian case under strong and very strong interference conditions. They also introduced an achievable scheme based on instantaneous amplify-and-forward relaying.
In \cite{Ho}, the authors studied the $K$-user interference channel in the presence of an IR in two scenarios, where the transmitters and the receivers are aware and not aware of the existence of the IR. It is shown that in both cases, the IR can enlarge the rate region and increase user fairness.
In \cite{Baik}, the authors studied general networks in the presence of IR and derived cut-set bounds for two cases of the IR having or not having its own message and showed that the proposed bounds are tight in some cases.
In \cite{Kramer}, it is proved that the networks with IR can be considered as a channel with in-block memory. Then, a cut-set bound was characterized, which generalizes existing cut-set bounds.

From the DoF perspective, it is well known that the DoF of frequency or time-selective $K$-user interference channel is $\frac{K}{2}$ \cite{Cadambe1}.
The sum DoF of a two-user interference channel assisted by an IR, with the same number of antennas for all nodes, was studied in \cite{Lee}, and it was proved that the DoF of $\frac{3}{2}$ can be achieved.
The DoF of the $M$ antenna three-user interference channel assisted by an IR was studied in \cite{Qiang}, and it was shown that the DoF of $2M$ is achievable.
The DoF of the two-way $K$-user IR-aided interference channel, when the IR is equipped with $2K$ antennas, was studied in \cite{Cheng}. It was demonstrated that the DoF of $K$ can be achieved.
The DoF of two-user interference channel in the presence of IR, when there is an arbitrary number of IR transmit and receive antennas, was studied in \cite{Liu}. An inner and two outer bounds were obtained.
For the $K$-user interference channel assisted by an IR, where the IR can only instantaneously amplify and forward the received signal in the current channel use, with the same number of antennas at all nodes, an achievable scheme and an outer bound were proposed in \cite{Azari}. Though the DoF in some special cases, where $K=2$ or the number of IRs is $K(K-1)$, was derived, a general achievable DoF was not obtained.

The goal of this paper is to study the achievable DoFs for the frequency-selective $K$-user interference channel in the presence of an IR.
To the best of our knowledge, although the DoF of two and three-user interference channels and the case where the number of IRs is $K(K-1)$ have been studied, achievable DoFs for the frequency-selective $K$-user interference channel (where the symbol extensions are in frequency domain) in the presence of a multi-input multi-output (MIMO) IR has not been characterized. Our contributions are as follows:
\begin{itemize}
\item
We prove an achievable DoF for the $K$-user interference channel in the presence of a MIMO IR with $Q$ receive antennas and $W$ transmit antennas, whose antennas can coordinate with each other, i.e., each transmit antenna has access to all receive antennas. For this purpose, we propose an interference alignment-based coding scheme, in which we divide the receivers into two groups, called clean and dirty receivers. We design the beamforming vectors such that some of the message symbols corresponding to the clean receivers can be demultiplexed at the IR while maintaining the basic properties of the interference alignment scheme such as aligning the interference vectors into smaller subspace at the receivers. By demultiplexing, we mean that the IR only separates some of the message symbols using linear operations, without removing the additive noise. Then the IR utilizes the demultiplexed symbols for interference cancellation at the clean receivers. Our proposed scheme increases the DoF for $W>\frac{K}{2}$ compared to the case without IR. Also, we show that if the number of IR antennas exceeds a finite threshold, the maximum DoF of $K$ can be achieved, and we characterize this threshold.
\item
Moreover, we derive the achievable DoF for a special kind of IR, for which the IR has the same number of receive and transmit antennas and the antennas cannot have coordination with each other, i.e., the $i$-th transmit antenna has access to the $i$-th receive antenna only. We extend the coding scheme for this case and derive an achievable DoF. Similar to the coordinated IR, we show that by considering the number of IR antennas more than a finite threshold, the maximum DoF of $K$ can be achieved. Our derivations show that the achievable DoF decreases considerably compared with the coordinated IR.
\end{itemize}

The paper is organized as follows.
In Section \ref{section2}, we present the system model.
In Sections \ref{sa} and \ref{sb}, we discuss our main results for the coordinated and non-coordinated IRs, respectively. In Section \ref{section4}, we present some numerical results to evaluate our proposed schemes. Finally, in Section \ref{section5}, we conclude the paper.

\textbf{Notations:}
Bold letters demonstrate matrices. Calligraphic upper case letters denotes sets and vector spaces. $\mathbb{R}$ is the set of real numbers. For a set $\cal A$, $|{\cal A}|$ indicates the cardinality of $\cal A$. ${\bf V}^{T}$ and ${\bf V}^{H}$ are the transpose and Hermitian  of matrix $\bf V$, respectively. ${\rm{diag} (a_1,...,a_m)}$ denotes a diagonal matrix with diagonal elements $a_1,...,a_m$.  The function $f(\rho)$ is $o(\log(\rho))$, if
\begin{equation*}
\mathop {\lim }\limits_{\rho  \to \infty } \frac{{|f(\rho )|}}{{\log (\rho )}} = 0.
\end{equation*}
Sequence $a(n)$ goes to infinity with $O(g(n))$, if
\begin{equation*}
0<\mathop {\lim }\limits_{n \to \infty } \frac{{\left| {a(n)} \right|}}{{|g(n)|}} < \infty .
\end{equation*}
$\mathbb{N}$ is the set of natural numbers and $\mathbb{W}$ is the set of non-negative integers.

\section{System Model and Preliminaries}
\label{section2}

\subsection{System model}

\begin{figure}
\centering
\includegraphics[width=15cm]{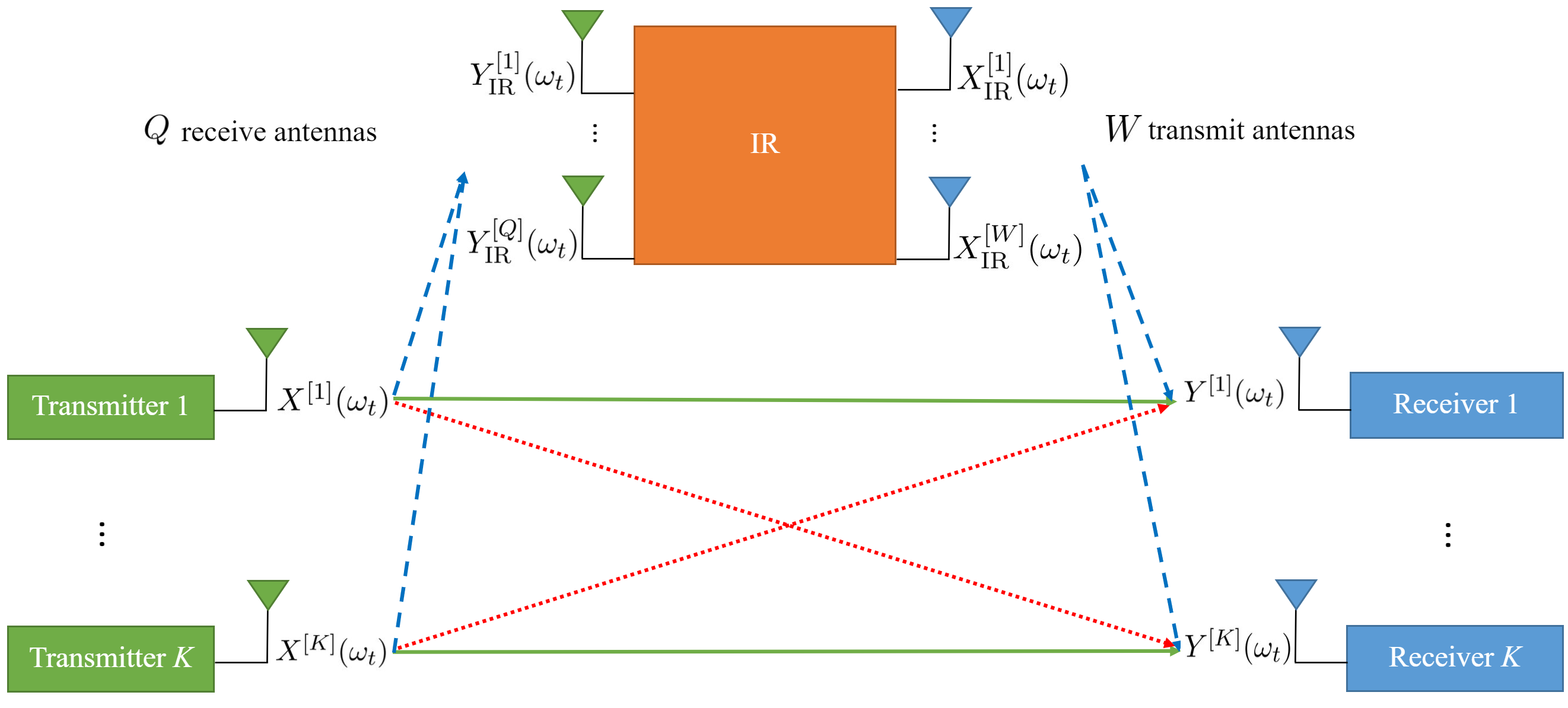}
\caption{IR-assisted $K$-user interference channel. The IR has $Q$ transmit antennas and $W$ receive antennas. Direct links are shown by solid arrows, cross-links are shown by  dotted arrows, and links between IRS and transmitters or receivers are shown by dashed arrows.}
 \label{illustration}
\end{figure}

We consider a $K$-user interference channel with an IR, in which $K$ single-antenna transmitters send their messages to $K$ single-antenna receivers. In this system, $i$-th transmitter sends the message $w^{[i]}\in{\cal W}^{[i]}=\left\{ {1,...,\left\lfloor {{2^{T{r_i}}}} \right\rfloor } \right\}$ to the $i$-th receiver, where $r_{i}$ is the transmission rate corresponding to the $i$-th transmitter and $T$ is the number of channel uses (in this paper each channel use corresponds to each frequency slot and all transmissions are in the same time cycle). We assume an IR with $Q$ receive antennas and $W$ transmit antennas. Fig. \ref{illustration} shows the system model. 

We consider a frequency-selective channel. Due to the instantaneity of the IR, it can process the signals received from all frequency slots in the current time cycle and transmit signals in different frequency slots in the same time cycle, which affects the received signals at the receivers in all frequency slots.
The received signal at the $j$-th receiver in the $t$-th frequency slot ${\omega _t}$ is shown by $Y^{[j]}({\omega _t})$ and is presented as follows\footnote{Note that in the general case, the IR transmitted signal is a function of the received signal in the past time cycles in addition to the current time cycle, but in this paper, we study achievability schemes, in which the signals of past time cycles are not needed and transmissions in different frequency slots are at the same time cycle.}:
\begin{equation}
{Y^{[j]}}({\omega _t}) = \sum\limits_{i = 1}^K {{H^{[ji]}}({\omega _t}){X^{[i]}}({\omega _t})}  + \sum\limits_{u = 1}^W {{H_{\rm IR-R}^{[ju]}}({\omega _t}){X_{\rm IR}^{[{u}]}}({\omega _t})}  + {Z^{[j]}}({\omega _t}),
\label{model receiver}
\end{equation}
where $X^{[i]}({\omega _t})$ is the signal of the $i$-th transmitter, ${H^{[ji]}}({\omega _t})$ is the channel coefficient between the $i$-th transmitter and the $j$-th receiver, ${{X_{\rm IR}^{[{u}]}}({\omega _t})}$ is the transmitted signal of the $u$-th IR transmit antenna, ${{H_{\rm IR-R}^{[ju]}}({\omega _t})}$ is the channel coefficient between the $u$-th  IR transmit antenna and the $j$-th receiver,  and ${Z^{[j]}}({\omega _t})$ is the additive white Gaussian noise (AWGN) at the $j$-th receiver  in the $t$-th frequency slot ${\omega _t}$, where $t \in \{1,2,...,T\}$. We assume perfect self-interference cancellation at the IR, thus, the received signal at the $q$-th IR receive antenna in the $t$-th frequency slot, which is shown by $Y_{\rm IR}^{[{q}]}({\omega _t})$, is given as follows:
\begin{equation}
{Y_{\rm IR}^{[{q}]}}({\omega _t}) = \sum\limits_{i = 1}^K {{H_{\rm T-IR}^{[{q}i]}}({\omega _t}){X^{[i]}}({\omega _t})}  + {Z_{\rm IR}^{[{q}]}}({\omega _t}),
\label{model relay}
\end{equation}
where ${H_{\rm T-IR}^{[{q}i]}}({\omega _t})$ is the channel coefficient from the $i$-th transmitter to the $q$-th IR receive antenna ($q \in \{1,...,Q\}$) and ${Z_{\rm IR}^{[{q}]}}({\omega _t})$ is AWGN at the $q$-th IR receive antenna in the $t$-th frequency slot. We assume that the perfect channel state information for all frequency slots is available at all nodes\footnote{This ideal assumption is vastly considered in the literature \cite{Cadambe1,Cadambe2}. The Noisy channel state information will be an interesting subject of the future work.}. We consider two types of IR: 1) a MIMO IR whose antennas can have coordination with each other called MIMO  coordinated IR (C-IR), 2) an IR with no coordination among its antennas, because the $u$-th transmit antenna only has access to the $u$-th receive antenna  ($W=Q$). We call this model non-coordinated IR (NC-IR). At each time cycle, for the MIMO C-IR, we have:
\begin{equation}
{X_{\rm IR}^{[{u}]}}({\omega _t}) = {f^{[u,{\omega _t}]}}({Y_{\rm IR}^{[{1}]}}({\omega _1}),...,{Y_{\rm IR}^{[{1}]}}({\omega _T}),...,{Y_{\rm IR}^{[{Q}]}}({\omega _1}),...,{Y_{\rm IR}^{[{Q}]}}({\omega _T})),
\label{func relay}
\end{equation}
where ${f^{[u,{\omega _t}]}}$ indicates the  encoding function of the IR for the $u$-th transmit antenna at the $t$-th frequency slot $\omega_t$. 
For the NC-IR, we have:
\begin{equation}
{X_{\rm IR}^{[{u}]}}({\omega _t}) = {f^{[u,{\omega _t}]}}({Y_{\rm IR}^{[{u}]}}({\omega _1}),...,{Y_{\rm IR}^{[{u}]}}({\omega _T})),u \in \{1,...,Q\}.
\label{func relay2}
\end{equation}
We limit the functions ${f^{[u,{\omega _t}]}}$ to be linear.

Since we assume a frequency-selective 
$K$-user interference channel,  ${H^{[ji]}}({\omega _t})$, ${{H_{\rm IR-R}^{[ju]}}({\omega _t})}$ and ${H_{\rm T-IR}^{[{q}i]}}({\omega _t})$ are independent random variables for different values of $i,j,u,q$ and $\omega_t$ with a continuous cumulative probability distribution due to frequency selectivity of the channel. In the case of complex channel coefficients, their real and imaginary parts are independent random variables with  continuous cumulative probability distribution function (e.g., complex Gaussian random variable). 

We assume that all  transmitters can send a signal with a maximum average power  $\rho $, i.e., $\frac{1}{T}\sum\limits_{t = 1}^T {{{\left| {{X^{[i]}}({\omega _t})} \right|}^2}}$  $\leqslant \rho ,\forall i \in \{ 1,...,K\} $. We say the rate vector ${{\bold r}} = ({r_1},...,{r_K})$ is achievable if $\mathop {\lim }\limits_{T \to \infty } \Pr \left\{ {\bigcap\limits_i {\{ {{\hat w}^{[i]}} \ne {w^{[i]}}\} } } \right\} = 0$, where ${{\hat w}^{[i]}}$ is the estimated message at the $i$-th receiver. In addition, ${\cal C}(\rho)$ indicates the closure of all achievable rate vectors ${{\bold r}} = ({r_1},...,{r_K})$.

\subsection{Preliminaries}

In the following, we introduce some definitions, which have been used throughout the paper. 

\textbf{Degrees of freedom (DoF)}:
Similar to \cite{Cadambe1}, we define the DoF region ${\cal D}$ for a $K$-user interference channel as follows:
\begin{equation*}
{{\cal D}} = \left\{ {({d_1},...,{d_K}) \in {\mathbb{R}}_ + ^K:\forall ({w_1},...,{w_K}) \in {\mathbb{R}}_ + ^K,} \right.
\end{equation*}
\begin{equation}
\left. {{w_1}{d_1} + ... + {w_K}{d_K} \le \mathop {\lim \sup }\limits_{\rho  \to \infty } \left(\frac{1}{\log(\rho)} {\mathop {\sup }\limits_{{{\bold r}}(\rho ) \in {{\cal C}}(\rho )} \left( {{w_1}{r_1} + ... + {w_K}{r_K}} \right)} \right)} \right\}.
\end{equation}

\textbf{Span}:
$\textrm{span}({\bold V})$ denotes the space spanned by the column vectors of the matrix $\bold V$, which is equal to ${\rm rank}({\bf V})$.

\textbf{Dimension}: We define the number of dimensions of $\textrm{span}({\bold V})$ as the dimension of $\bold V$  and show it by $d({\bold V})$.

\textbf{Normalized asymptotic dimension}:
 We will see in our analysis that for a given $K,Q$, and $W$, the dimension of the beamforming matrices and the vector spaces will have an order  of $O(n^{l}),l,n \in \mathbb{N}$. For a  matrix $\bold V$, we define the normalized asymptotic dimension (${D_N}$)  as follows:
\begin{equation}
{D_N}({\bold V}) = \mathop {\lim }\limits_{n \to \infty } \frac{{d({\bold V})}}{{{n^l}}},
\label{DN}
\end{equation}
where $l$ is the maximum integer number that satisfies $\mathop {\lim }\limits_{n \to \infty } \frac{{d(\bf{V})}}{{{n^l}}} < \infty$.

These definitions are also used for a vector space $\cal{A}$, therefore, $d({\cal A})$ indicates the dimension of $\cal A$ and $D_N({\cal A})$ indicates the normalized asymptotic  dimension of $\cal A$.

\section{Frequency-Selective $K$-User Interference Channel in the Presence of MIMO C-IR}
\label{sa}
In this section, we present an achievability theorem for the DoF of the frequency-selective $K$-user interference channel with a MIMO C-IR:
\begin{theorem}
For a frequency-selective $K$-user interference channel with a MIMO C-IR, where $\max \{ W,Q\} \le K$, the following DoF is achievable:
\begin{equation}
{\rm DoF}= \max \left\{ {\frac{K}{2} + \max \left\{ {0,K\frac{{\frac{W}{K} - \frac{1}{2}}}{{1 + 2\left\lceil {\frac{W}{Q}} \right\rceil }}} \right\},\min \left\{ {Q,W} \right\}} \right\}.
\label{DoF1}
\end{equation}
\label{theorem1}

We can see from (\ref{DoF1}) that when $\frac{W}{K} > \frac{1}{2}$, the DoF always increases over $\frac{K}{2}$, i.e., the DoF in the absence of an IR.
\end{theorem}

\begin{IEEEproof}
We rewrite (\ref{model receiver}) and (\ref{model relay}) into the following vector form:
\begin{equation}
{\bold{Y}^{[j]}} = \sum\limits_{i = 1}^K {{\bold{H}^{[ji]}}{\bold{X}^{[i]}}}  + \sum\limits_{u = 1}^W {{\bold{H}_{\rm IR-R}^{[j{u}]}}{\bold{X}_{\rm IR}^{[{u}]}}}  + {\bold{Z}^{[j]}},
\label{output}
\end{equation}
\begin{equation}
{\bold{Y}_{\rm IR}^{[{q}]}} = \sum\limits_{i = 1}^K {{\bold{H}_{\rm T-IR}^{[{q}i]}}{\bold{X}^{[i]}}}  + {\bold{Z}_{\rm IR}^{[{q}]}},
\end{equation}
where $\bold{X}^{[i]}$ is a $T \times 1$ column vector including channel inputs ${X}^{[i]}(\omega_t)$, i.e.,
\begin{equation*}
{{\bf{X}}^{[i]}} = {\left[ {\begin{array}{*{20}{c}}
{{X^{[i]}}({\omega _1})}&{{X^{[i]}}({\omega _2})}& \cdots &{{X^{[i]}}({\omega _T})}
\end{array}} \right]^T}.
\end{equation*}
$\bold{Y}^{[i]}$, $\bold{Y}_{\rm IR}^{[{q}]}$, $\bold{X}_{\rm IR}^{[u]}$, $\bold{Z}^{[j]}$ and $\bold{Z}_{\rm IR}^{[{q}]}$ are also defined in the similar way. ${\bold{H}^{[ji]}}$ is a diagonal matrix defined as follows:
\begin{equation*}
{{\bf{H}}^{[ji]}} = {\rm diag}\left( {{H^{[ji]}}({\omega _1}),...,{H^{[ji]}}({\omega _T})} \right).
\end{equation*}
${\bold{H}_{\rm IR-R}^{[j{u}]}}$ and ${\bold{H}_{\rm T-IR}^{[{q}i]}}$ are also defined similarly. We will prove the achievability of the  first term $\frac{K}{2} + \max \left\{ {0,K\frac{{\frac{W}{K} - \frac{1}{2}}}{{1 + 2\left\lceil {\frac{W}{Q}} \right\rceil }}} \right\}$ in (\ref{DoF1}) in the following. The proof of second term, i.e., $\min\{Q,W\}$ is provided  in Appendix \ref{appendix0}.

We present this proof in 6 steps. In Step 1, we divide the transmitters and the receivers into two groups (clean and dirty). In Step 2, some message streams are considered to have the capability of being demultiplexed at the MIMO C-IR, thus, the MIMO C-IR can use them for interference cancellation in clean receivers. After interference cancellation, the equivalent channel coefficients will be derived for other receivers (dirty receivers). In Step 3, we introduce the interference alignment equations such that the assumption of the previous step (demultiplexing of some message streams) and interference alignment for each receiver and MIMO C-IR receive antenna are satisfied. In Step 4, we present the beamforming design for each symbol stream. In Step 5, we analyze the satisfaction of the interference alignment equations at each receiver and MIMO C-IR receive antenna. Finally, in Step 6, we derive the achieved DoF, presented in the first term of (\ref{DoF1}).

\vspace{30pt}

\textbf{Step 1: Partitioning the  transmitters and receivers}

We divide the transmitters in two partitions. For the transmitters $ i \in \{1,...,W\}$, we provide two sets of symbol streams ${{\bar {\bold{x}}}^{[i]}}$ and ${{\tilde {\bold{x}}}^{[i]}}$ (each element of the vectors ${{\bar {\bold{x}}}^{[i]}}$ and ${{\tilde {\bold{x}}}^{[i]}}$ are the extended symbols).  The matrices ${{\bar{\bold{V}}}^{[i]}}$ and  ${{\tilde{\bold{V}}}^{[i]}}$ are the beamforming matrices, whose columns are the beamforming vectors corresponding to the elements of ${{\bar {\bold{x}}}^{[i]}}$ and ${{\tilde{\bold{x}}}^{[i]}}$,
 respectively, we can write:
\begin{equation}
{\bold{X}^{[i]}} = {{\bar{\bold{V}}}^{[i]}}{{\bar{\bold{x}}}^{[i]}} + {{\tilde{\bold{V}}}^{[i]}}{{\tilde{\bold{x}}}^{[i]}},i\in \{1,...,W\}.
\label{partition1}
\end{equation}
For the transmitters $i\in \{W+1,...,K\}$,  we only provide one set of extended symbols ${{\bar {\bold{x}}}^{[i]}}$, and ${{\bar{\bold{V}}}^{[i]}}$ is the beamforming matrix for the symbols ${{\bar {\bold{x}}}^{[i]}}$. So, we have:
\begin{equation}
{\bold{X}^{[i]}} = {{\bar{\bold{V}}}^{[i]}}{{\bar {\bold{x}}}^{[i]}},i \in \{W+1,...,K\}.
\label{partition2}
\end{equation}
Note that the matrices ${{\tilde{\bold{V}}}^{[i]}}$ and ${{\bar{\bold{V}}}^{[i]}}$ have $T$ rows, because we have $T$ frquency slots. The dimension of ${{\bar {\bold{x}}}^{[i]}}$ and ${{\tilde {\bold{x}}}^{[i]}}$, and the number of columns of ${{\bar{\bold{V}}}^{[i]}}$ and ${{\tilde{\bold{V}}}^{[i]}}$ will be determined in the next steps.

In the next following steps, we design the beamforming vectors ${{\tilde{\bold{V}}}^{[i]}}$ and ${{\bar{\bold{V}}}^{[i]}}$ such that the extended symbols ${{\tilde {\bold{x}}}^{[i]}}$ can be demultiplexed at the MIMO C-IR. By demultiplexing, we mean that the MIMO C-IR can separate each symbol of message streams $\tilde{\bold x}^{[i]}$ using zero forcing without decoding it. The symbol stream ${{\bar {\bold{x}}}^{[i]}}$ act as interference signals and their beamforming vectors align into a smaller subspace.

We also divide the receivers into clean and dirty sets. In the next steps, the transmitted signal by the MIMO C-IR is designed such that the interference induced by the symbols ${{\tilde {\bold{x}}}^{[i]}}$ will be removed at the receivers $j \in \{1,...,W\}$ called clean receivers, but this interference will remain at the receivers $j\in \{W+1,...,K\}$ called dirty receivers.

\vspace{10pt}

\textbf{Step 2: Interference cancellation at clean receivers and equivalent channel for dirty receivers}

We design the beamforming vectors ${{\tilde{\bold{V}}}^{[i]}}$ and  ${{\bar{\bold{V}}}^{[i]}}$ such that the interference induced by the symbols   ${{\tilde {\bold{x}}}^{[i]}}$  will be removed at clean receivers. We denote this  interference as ${\tilde {\bold{I}}^{[j]}} $, which is written as follows:
\begin{equation}
{\tilde {\bold{I}}^{[j]}} = \sum\limits_{i \in \{ 1,...,W\} ,i \ne j} {{{\bold{H}}^{[ji]}}{{\tilde {\bold{V}}}^{[i]}}{{\tilde {\bold{x}}}^{[i]}}} ,j \in \{ 1,...,W\},
\end{equation}
The MIMO C-IR can demultiplex the streams corresponding to $\tilde{\bf x}^{[i]}$ (will be shown in Steps 3-5), which is only contaminated by an additive noise, i.e., it will separate them into the form of ${\hat{\tilde{\bf x}}}^{[i]}=\tilde{\bf x}^{[i]}+\tilde{\bf z}^{[i]}$.
Thus, for the interference cancellation, the MIMO C-IR designs its transmitted signal such that:
\begin{equation}
\scalebox{.96}[1]{$\sum\limits_{u \in \{ 1,...,W\} } {{{\bf{H}}_{\rm IR-R}^{[j{u}]}}{{\bf{X}}_{\rm IR}^{[{u}]}}} =  - \sum\limits_{i \in \{ 1,...,W\} ,i \ne j} {{{\bf{H}}^{[ji]}}{{\tilde {\bf{V}}}^{[i]}} {\hat{\tilde{\bf x}}}^{[i]} } =  - \sum\limits_{i \in \{ 1,...,W\} ,i \ne j} {{{\bf{H}}^{[ji]}}{{\tilde {\bf{V}}}^{[i]}}\left( {{{\tilde {\bf{x}}}^{[i]}} + {{\tilde {\bf{z}}}^{[i]}}} \right)}  =  - {\tilde {\bf{I}}^{[j]}} + {\tilde {\bf{Z}}^{[j]}},$}
\label{interference cancellation}
\end{equation}
where
\begin{equation*}
{\tilde {\bf{Z}}^{[j]}}= - \sum\limits_{i \in \{ 1,...,W\} ,i \ne j} {{{\bf{H}}^{[ji]}}{{\tilde {\bf{V}}}^{[i]}}{{\tilde {\bf{z}}}^{[i]}}}  .
\end{equation*}
The vector equation (\ref{interference cancellation}) generates a linear set of  equations, an equation for each element of ${\bold{X}_{\rm IR}^{[{u}]}}$, which can be written for the $t$-th element  as:
\begin{equation}
\sum\limits_{u \in \{ 1,...,W\} } {{H_{\rm IR-R}^{[ju]}}({\omega _t}){X_{\rm IR}^{[{u}]}}} ({\omega _t}) =  - {{\tilde I}^{[j]}}({\omega _t})+\tilde{ Z}^{[j]}(\omega_t),\quad\forall j\in \{1,...,W\},\quad \forall t\in\{1,...,T\},
\label{interference cancellation2}
\end{equation}
which is a linear set of equations with $W$ variables for each $\omega_t$. This set of equations is almost surely solvable,  since the coefficients of the linear equations are drawn independently from a continuous cumulative probability distribution, thus, the determinant of the matrix of linear equations will be a non-zero polynomial in terms of independent random variables and by \cite[Lemma 1]{me}, it will be non-zero with probability equal to $1$.
Applying (\ref{interference cancellation2}), the interference cancellation will be done.
So, for each $\omega_t$, we will have:
\begin{equation}
{X_{\rm IR}^{[{u}]}}({\omega _t}) = \sum\limits_{j \in \{ 1,...,W\} } {H_{\rm inv}^{[j{u}]}({\omega _t})( - {{\tilde I}^{[j]}}({\omega _t}) ({\omega _t})+\tilde{ Z}^{[j]}(\omega_t))},
\label{interference cancellation3}
\end{equation}
where $H_{\rm inv}^{[j{u}]}({\omega _t})$, the factor of $ - {{\tilde I}^{[j]}}({\omega _t})+\tilde{ Z}^{[j]}(\omega_t)$ in (\ref{interference cancellation3}), is a function of ${H_{\rm IR-R}^{[j'{{u'}}]}}({\omega _t}),u',j'\in\{1,...,W\}$ obtained by  solving equations  
(\ref{interference cancellation2}).
We can write equations (\ref{interference cancellation3}) in the vector form as follows:
\begin{align}
{\bold{X}_{\rm IR}^{[{u}]}} &= \sum\limits_{j \in \{ 1,...,W\} } {\bold{H}_{\rm inv}^{[j{u}]}( - {{\tilde {\bold{I}}}^{[j]}}+\tilde{\bf Z}^{[j]})}\\ 
& = \sum\limits_{j \in \{ 1,...,W\} } {\sum\limits_{i \in \{ 1,...,W\} ,i \ne j} {-\bold{H}_{\rm inv}^{[j{u}]}{\bold{H}^{[ji]}}{{\tilde {\bold{V}}}^{[i]}}{{\tilde {\bold{x}}}^{[i]}} + } } \sum\limits_{j \in \{ 1,...,W\} } {\bold{H}_{\rm inv}^{[j{u}]}{{\tilde {\bold{Z}}}^{[j]}}} ,
\label{refrence eq}
\end{align}
where $\bold{H}_{\rm inv}^{[j{u}]}$ is a diagonal matrix as follows:
\begin{equation*}
{\bf{H}}_{\rm inv}^{[j{u}]} = {\rm diag}\left( {H_{\rm inv}^{[j{u}]}({\omega _1}),...,H_{\rm inv}^{[j{u}]}({\omega _T})} \right).
\end{equation*}
We highlight two properties of $\bold{H}_{\rm inv}^{[j{u}]}$:
\begin{itemize}
\item
Like $\bold{H}^{[ji]}$, diagonal elements ${H}_{\rm inv}^{[j{u}]}(\omega_t)$ are independent random variables for different $t\in \{1,...,T\}$, because the channel coefficients are independent random variables for each $t\in\{1,...,T\}$.

\item
Each diagonal element ${H}_{\rm inv}^{[j{u}]}(\omega_t)$ is a fractional polynomial constructed by ${H}_{\rm IR-R}^{[j'{u'}]}(\omega_t),j',u'\in\{1,...,W\}$.
A fractional polynomial is the ratio of polynomial $P_1(\cdot)$ to non-zero polynomial $P_2(\cdot)$.
\end{itemize}

Although we cancel the interference ${\tilde {\bold{I}}^{[j]}} $ at clean receivers, this interference will remain at dirty receivers with new equivalent channel coefficients.
Now, we derive the new channel coefficients for ${{\tilde {\bold{V}}}^{[i]}}{{\tilde {\bold{x}}}^{[i]}},\forall i\in \{1,...,W\}$ at dirty receivers $j\in \{W+1,...,K\}$. By combining (\ref{output}), (\ref{partition1}) and (\ref{partition2}), we have: 
\begin{align}
{{\bf{Y}}^{[j]}}& = \sum\limits_{i \in \{ 1,...,K\} } {{{\bf{H}}^{[ji]}}{{\bar {\bf{V}} }^{[i]}}{{\bar {\bf{x}} }^{[i]}}}  + \sum\limits_{i \in \{ 1,...,W\} } {{{\bf{H}}^{[ji]}}{{\tilde {\bf{V}}}^{[i]}}{{\tilde {\bf{x}}}^{[i]}}}  + \sum\limits_{u \in \{ 1,...,W\} } {{{\bf{H}}_{\rm IR-R}^{[j{u}]}}{{\bf{X}}_{\rm IR}^{[{u}]}}}  + {{\bf{Z}}^{[j]}}\\
 &= \sum\limits_{i \in \{ 1,...,K\} } {{{\bf{H}}^{[ji]}}{{\bar {\bf{V}} }^{[i]}}{{\bar {\bf{x}} }^{[i]}}}  + \sum\limits_{i \in \{ 1,...,W\} } {{{\bf{H}}^{[ji]}}{{\tilde {\bf{V}}}^{[i]}}{{\tilde {\bf{x}}}^{[i]}}}  + \sum\limits_{u,d,i \in \{ 1,...,W\} ,i \ne d} {{{\bf{H}}_{\rm IR-R}^{[j{u}]}}{\bf{H}}_{\rm inv}^{[d{u}]}{{\bf{H}}^{[di]}}{{\tilde {\bf{V}} }^{[i]}}{{\tilde {\bf{x}} }^{[i]}}}  + {{{\bf{\tilde {\tilde{ Z}}}}}^{[j]}},
\label{output equivalent}
\end{align}
where (\ref{output equivalent}) follows from (\ref{refrence eq}) and:
\begin{equation*}
{{{\bf{\tilde {\tilde {Z}}}}}^{[j]}} = \sum\limits_{u,d} {{{\bf{H}}_{\rm IR-R}^{[j{u}]}}{\bf{H}}_{\rm inv}^{[d{u}]}} {{\tilde{\bf{Z}}}^{[d]}} + {{\bf{Z}}^{[j]}}.
\end{equation*}
(\ref{output equivalent}) can be rewritten as:
\begin{equation}
{{\bf{Y}}^{[j]}} = \sum\limits_{i \in \{ 1,...,K\} } {{{\bf{H}}^{[ji]}}{{\bar {\bf{V}} }^{[i]}}{{\bar {\bf{x}} }^{[i]}}}    + \sum\limits_{i \in \{ 1,...,W\} } {{{{\tilde{\bf H}}}^{[ji]}}{{\tilde {\bf{V}}}^{[i]}}{{\tilde {\bf{x}}}^{[i]}}}  + {{{\bf{\tilde {\tilde {Z}}}}}^{[j]}},
\label{receivers}
\end{equation}
\begin{equation}
{{{{\tilde {\bf H}}}}^{[ji]}} = {{\mathbf{H}}^{[ji]}} + \sum\limits_{u,d \in \{ 1,...,W\} ,d \ne i} {{\mathbf{H}}_{{\rm{IR-R}}}^{[ju]}{\mathbf{H}}_{{\rm{inv}}}^{[du]}{{\mathbf{H}}^{[di]}}} ,i\in\{1,...,W\} ,
\label{tilde h}
\end{equation}
where ${{{\tilde{\bf H}}}^{[ji]}}$ is the equivalent channel coefficient matrix from the transmitter $i \in \{1,...,W\}$ to the receiver $j \in \{W+1,...,K\}$ (dirty receivers) for ${{\tilde {\bf{V}}}^{[i]}}{{\tilde {\bf{x}}}^{[i]}}$.
By (\ref{tilde h}), we can see that ${{{\tilde{\bf H}}}^{[ji]}}$ has the following properties:
\begin{itemize}
\item
${{{\tilde{\bf H}}}^{[ji]}}$ is a diagonal matrix.

\item
${{\tilde{\bf{ H}}}^{[ji]}} = {{\bf{H}}^{[ji]}},\forall j \in \{ 1,...,W\} $.

\item
For $j \in \{W+1,...,K\}$, its $t$-th diagonal element has the following form:
\begin{equation*}
\scalebox{.92}[1]{${{\tilde H}^{[ji]}}({\omega _t}) = \sum\limits_{u,i',j' \in \{ 1,...W\} ,i' \ne j'} {{H_{\rm IR-R}^{[ju]}}({\omega _t}){H^{[j'i']}}({\omega _t}){P^{[ui'j']}}(\{{H_{\rm IR-R}^{[m{e}]}}({\omega _t}):m,e \in \{ 1,...,W\}\} )}  + {H^{[ji]}}({\omega _t}),$}
\end{equation*} 
where $P^{[ui'j']}({\cal S})$ indicates a fractional polynomial constructed from the variables $s \in {\cal S}$.
\end{itemize}

\vspace{10pt}

\textbf{Step 3: Interference alignment}

In this step, we determine the interference alignment equations in clean and dirty receivers and MIMO C-IR receive antennas. In our interference alignment scheme, we align the subspace of interference of each user into a bigger subspace with an equal normalized asymptotic dimension. Note that for a matrix $\bold V$ and $\bold V'$, we can have the following relations simultaneously, $d({\bold V})>d({\bold V}')$, $D_N({\bold V})=D_N({\bold V}')$, e.g., $d({\bold V})={(n+1)}^l>d({\bold V}')=n^l$, $D_N({\bold V})=D_N({\bold V}')=1$. We begin with clean receivers.

\vspace{10pt}

\textit{1) Interference alignment at clean receivers:}

Consider a clean receiver $j \in \{1,...,W\}$, for each $i\in \{1,...,K\},i\ne j$, we must have:
\begin{equation}
\textrm{span}\left( {{\bold{H}^{[ji]}}{{\bar {\bold V}}^{[i]}}} \right) \subseteq {{\bar{\cal A}}_j},
\label{IA1}
\end{equation}
where ${\bar{\cal A}_j}$ is considered as a subspace that encompass all interference at the $j$-th receiver induced by $\bar {\bf x}^{[i]},i\in\{1,...,K\},i\ne j$, for which we have:
\begin{equation}
\mathop {\max }\limits_{i\in \{1,...,K\},i \ne j} {D_N}\left(\textrm{span}\left({\bold{H}^{[ji]}}{{\bar {\bold V}}^{[i]}}\right)\right) = {D_N}({\bar{\cal A}_j}),
\label{IA2}
\end{equation}
which implies that the normalized asymptotic dimension of ${\bar{\cal A}_j}$ is equal to the maximum asymptotic dimension of $\textrm{span}\left({\bold{H}^{[ji]}}{{\bar {\bold V}}^{[i]}}\right)$, for $\forall i\ne j$.
Also, we define the message subspaces as:
\begin{equation*}
{\bar {\cal C}_j} = \textrm{span}\left( {{{\bold H}^{[jj]}}{{\bar {\bold V}}^{[j]}}} \right),
\end{equation*}
\begin{equation*}
{\tilde {\cal C}_j} = \textrm{span}\left( {{{\tilde{\bold H}}^{[jj]}}{{\tilde {\bold V}}^{[j]}}} \right).
\end{equation*}
and we require ${\bar {\cal C}_j}$, ${\tilde {\cal C}_j}$, and ${\bar {\cal A}_j}$ to be full rank and linearly independent, thus, we can ensure the decodability of the message streams $\tilde {\bold x}^{[j]}$ and $\bar {\bold x}^{[j]}$ by zero forcing at the $j$-th receiver.

\vspace{10pt}

\textit{2) Interference alignment at dirty receivers:}

Consider a dirty receiver $j \in \{W+1,...,K\}$. 
Here, we have two interference subspaces at each receiver $j$, the interference induced by $\bar{\bf x}^{[i]}$ aligns in subspace $\bar{\cal A}_j$, while the interference induced by $\tilde{\bf x}^{[i]}$ aligns in subspace $\tilde{\cal A}_j$.
For each $i\in \{1,...,K\},i\ne j$, we must have:
\begin{equation}
\textrm{span}\left( {{\bold{H}^{[ji]}}{{\bar {\bold V}}^{[i]}}} \right) \subseteq {\bar{\cal A}_j},
\label{IA3}
\end{equation}
where ${\bar{\cal A}_j}$ is considered as a subspace, for which we have:
\begin{equation}
\mathop {\max }\limits_{i\in \{1,...,K\},i \ne j} {D_N}\left(\textrm{span}\left({\bold{H}^{[ji]}}{{\bar {\bold V}}^{[i]}}\right)\right) = {D_N}({\bar{\cal A}_j}),
\label{IA4}
\end{equation}
and for every $i\in \{1,...,W\}$, we must have:
\begin{equation}
\textrm{span}\left( {{\tilde{\bold{H}}^{[ji]}}{{\tilde {\bold V}}^{[i]}}} \right) \subseteq {\tilde{\cal A}_j},
\label{IA5}
\end{equation}
where ${\tilde{\cal A}_j}$ is considered as a subspace, for which we have:
\begin{equation}
\mathop {\max }\limits_{i \in \{1,...,W\}} {D_N}\left(\textrm{span}\left({\tilde{\bold{H}}^{[ji]}}{{\tilde {\bold V}}^{[i]}}\right)\right) = {D_N}({\tilde{\cal A}_j}).
\label{IA6}
\end{equation}

Also, we define the message subspace as:
\begin{equation*}
{\bar {\cal C}_j} = \textrm{span}\left( {{{\bold H}^{[jj]}}{{\bar {\bold V}}^{[j]}}} \right),
\end{equation*}
and we want ${\bar {\cal C}_j}$, ${\tilde {\cal A}_j}$ and ${\bar {\cal A}_j}$ to be full rank and linearly independent, hence, we can ensure the decodability of the message stream $\bar {\bold x}^{[j]}$ by zero forcing in the $j$-th receiver.

\vspace{10pt}

\textit{3) Interference alignment at the MIMO C-IR $q$-th  receive antenna:}

We assume that $W=QZ+P,0\le P<Q$, we divide the transmitters $i\in\{1,...,W\},$ into $Q$ distinct sets, the first $P$ sets include $Z+1$ transmitters and other $Q-P$ sets include $Z$ transmitters.
We name these sets ${{\cal B}}_q,q\in \{1,...,Q\}$. We design our interference alignment scheme such that symbol streams $\tilde{\bf x}^{[i]},i\in{\cal B}_q$ can be demultiplexed at the $q$-th receive antenna of the MIMO C-IR.
To this end, all the interference induced by the symbol streams $\bar {\bf x}^{[i]},i\in\{1,...,K\}$, must align into a limited subspace at each receive antenna of the MIMO C-IR.
Thus, at each receive antenna $q\in\{1,...,Q\}$, and for each $i \in \{1,...,K\}$, we must have:
\begin{equation}
\textrm{span}\left( {{\bold{H}_{\rm T-IR}^{[qi]}}{{\bar {\bold V}}^{[i]}}} \right) \subseteq {\bar{\cal A}_{r_q}},
\label{IA7}
\end{equation}
where ${\bar{\cal A}_{r_q}}$ is considered as a subspace, for which we have:
\begin{equation}
\mathop {\max }\limits_{i\in \{1,...,K\}} {D_N}\left(\textrm{span}\left({\bold{H}_{\rm T-IR}^{[qi]}}{{\bar {\bold V}}^{[i]}}\right)\right) = {D_N}({\bar{\cal A}_{r_q}}).
\label{IA8}
\end{equation}
In addition, at the $q$-th receive antenna of the MIMO C-IR, the interference induced by symbol streams $\tilde{\bf x}^{[i]},i \in \{1,...,W\},i  \notin  {{\cal B}}_q$, must align into a subspace named $\tilde{\cal A}_{r_q}$. Hence, for each $i \in \{1,...,W\},i  \notin  {{\cal B}}_q$, we must have:
\begin{equation}
\textrm{span}\left( {{\bold{H}_{\rm T-IR}^{[qi]}}{{\tilde {\bold V}}^{[i]}}} \right) \subseteq {\tilde{\cal A}_{r_q}},
\label{IA9}
\end{equation}
where ${\tilde{\cal A}_{r_q}}$ is considered as a subspace, for which we have:
\begin{equation}
\mathop {\max }\limits_{i \in \{1,...,W\},i  \notin  {{\cal B}}_q} {D_N}\left(\textrm{span}\left({\bold{H}_{\rm T-IR}^{[qi]}}{{\tilde {\bold V}}^{[i]}}\right)\right) = {D_N}({\tilde{\cal A}_{r_q}}).
\label{IA10}
\end{equation}

Also, we define ${\tilde {\cal C}_{i,r_q}},i\in {\cal B}_q$ as the message subspaces, which can be demultiplexed at the $q$-th MIMO C-IR receive antenna as follows:
\begin{equation*}
{\tilde {\cal C}_{i,r_q}} = \textrm{span}\left( {{{{\bold H}}_{\rm T-IR}^{[qi]}}{{\tilde {\bold V}}^{[i]}}} \right),i \in {\cal B}_q,
\end{equation*}
and we want ${\tilde {\cal C}_{i,r_q}}, \forall i\in {\cal B}_q$, $\bar {\cal A}_{r_q}$ and $\tilde {\cal A}_{r_q}$ to be full rank and linearly independent,  thus, we can make sure that the message streams $\tilde {\bold x}^{[i]},i\in {\cal B}_q$ can be demultiplexed at the $q$-th MIMO C-IR receive antenna  by zero forcing. Note that  $q$-th receive antenna of the MIMO C-IR demultiplexes the message streams $\tilde {\bold x}^{[i]},i\in {\cal B}_q$ without having coordination with other receive antennas. After each antenna demultiplexes its own message streams $\tilde {\bold x}^{[i]},i\in {\cal B}_q$, all of these message streams are passed to the MIMO C-IR transmit antennas, so the transmit antennas can have coordination with each other for interference cancellation at the clean receivers (as Eq. (\ref{refrence eq})). A simple illustration for the interference alignment scheme is shown in Fig. \ref{fig10} for $K=3$ and $W=2$. In Steps 4 and 5, we prove the existence of such beamforming vectors, message, and interference subspaces, which satisfies the previous interference alignment equations (\ref{IA1})-(\ref{IA10}) for clean and dirty receivers and the MIMO C-IR. In Step 6, we analyze the achieved DoF by these beamforming vectors design.

\begin{figure}
\centering
\includegraphics[width=13cm]{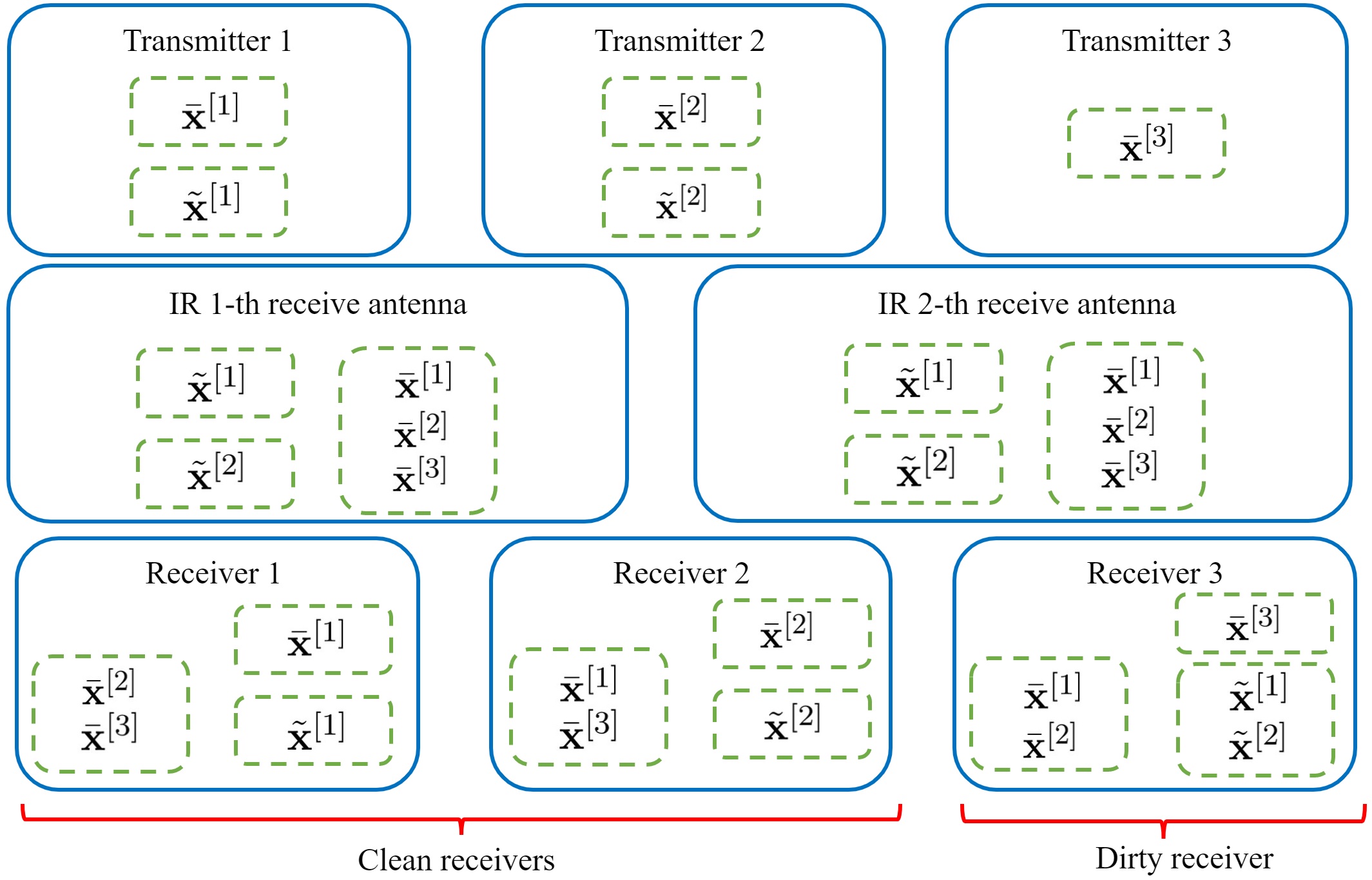}
\caption{Interference alignment scheme for $3$-user interference channel in the presence of MIMO C-IR with $2$ receive antennas. Subspaces corresponding to symbol streams in common dashed boxes align into a joint subspace at each node. We can see that the interference of the message streams $\tilde{\bf x}^{[1]}$ and $\tilde{\bf x}^{[2]}$ is canceled at clean receivers.}
 \label{fig10}
\end{figure}


\vspace{10pt}

\textbf{Step 4: Beamforming matrix design}

In this step, we design beamforming matrices such that the alignment equations (\ref{IA1})-(\ref{IA10}) are satisfied and all users' message streams are decodable.


\textit{1) Beamforming matrix design for $i\in \{1,...,W\}$}:

Consider the vector ${\bf w} = {\left[ {\begin{array}{*{20}{c}}
1&1& \cdots &1
\end{array}} \right]^H}$. We design the beamforming matrices ${{\bar{\bold{V}}}^{[i]}}$ and ${{\tilde{\bold{V}}}^{[i]}}$ as the following:
\begin{equation}
{\bar {\bf{V}} ^{[i]}} = \left\{ {\left[ {\prod\limits_{(i',j')\in\bar {\cal S}_1^{[i]}} {{{\left( {{{\bf{H}}^{[j'i']}}} \right)}^{{\alpha _{j'i'}}}}} } \right]\left[ {\prod\limits_{(i'',{q'})\in\bar {\cal S}_2^{[i]}} {{{\left( {{{\bf{H}}_{\rm T-IR}^{[{{q'}}i'']}}} \right)}^{{\gamma _{{{q'}}i''}}}}} } \right]{\bf w}:{\alpha _{j'i'}} \in \{ 1,...,n\} ,{\gamma _{{{q'}}i''}} \in \{ 1,...,sn\} } \right\},
\label{beam1}
\end{equation}
where we define:
\begin{equation}
\bar {\cal S}_1^{[i]} = \left\{ {(i',j')\left| {i',j' \in \{ 1,...,K\} ,i' \ne j'} \right.} \right\},
\label{set1}
\end{equation}
\begin{equation}
\bar {\cal S}_2^{[i]} = \left\{ {(i'',q')\left| {i'' \in \{ 1,...,K\} ,q' \in \{ 1,...,Q\} } \right.} \right\},
\label{set2}
\end{equation}
$n \in \mathbb{N}$ is an auxiliary variable which can go to infinity and $s$ is a parameter for controlling the dimension of ${{\bar{\bold{V}}}^{[i]}}$, i.e., $d({{\bar{\bold{V}}}^{[i]}})$. (\ref{beam1}) indicates that $\alpha_{j'i'}$  and $\gamma_{{q'}i''}$ can take any value from the sets $\{1,...,n\}$ and $\{1,...,sn\}$, therefore, the number of columns of  ${\bar {\bf{V}} ^{[i]}}$ will be $n^{K^2-K}{(sn)}^{QK}$. This notation means that the right-hand side of (\ref{beam1}) is the set of column vectors, which form the beamforming matrix ${\bar {\bf{V}} ^{[i]}}$\footnote{The order of these column vectors is not important. In particular, for each value of ${\alpha _{j'i'}}\in\{1,...,n\}$ and ${\gamma _{{{q'}}i''}}\in\{1,...,sn\}$, the vector $\left[ {\prod\limits_{(i',j')\in\bar {\cal S}_1^{[i]}} {{{\left( {{{\bf{H}}^{[j'i']}}} \right)}^{{\alpha _{j'i'}}}}} } \right]\left[ {\prod\limits_{(i'',{q'})\in\bar {\cal S}_2^{[i]}} {{{\left( {{{\bf{H}}_{\rm T-IR}^{[{{q'}}i'']}}} \right)}^{{\gamma _{{{q'}}i''}}}}} } \right]{\bf w}$ is a beamforming vector corresponding to one element of $\bar{\bf x}^{[i]}$. For more clarity, if we assume $n=s=1$, then the matrix ${\bar {\bf{V}} ^{[i]}}$ contains only one column vector $\left[ {\prod\limits_{(i',j')\in\bar {\cal S}_1^{[i]}} {{{\left( {{{\bf{H}}^{[j'i']}}} \right)}}} } \right]\left[ {\prod\limits_{(i'',{q'})\in\bar {\cal S}_2^{[i]}} {{{\left( {{{\bf{H}}_{\rm T-IR}^{[{{q'}}i'']}}} \right)}}} } \right]{\bf w}$.}.
For ${{\tilde{\bold{V}}}^{[i]}}$, we have:
\begin{equation*}
{{{\tilde{\bf{ V}}}}^{[i]}} = \left\{ {\left[ {\prod\limits_{(i',j')\in\tilde {\cal S}_1^{[i]}} {{{\left( {{{{\tilde{\bf H}}}^{[j'i']}}} \right)}^{{\alpha _{j'i'}}}}} } \right]\left[ {\prod\limits_{(i'',q')\in\tilde {\cal S}_2^{[i]}} {{{\left( {{{\bf{H}}_{\rm T-IR}^{[{{q'}}i'']}}} \right)}^{{\gamma _{{{q'}}i''}}}}} } \right]\left[ {\prod\limits_{(i''',q'')\in\tilde {\cal S}_3^{[i]}} {{{\left( {{\bold{T}^{[{{q''}}i''']}}} \right)}^{{\beta _{{{q''}}i'''}}}}} } \right]{\bf w}:} \right.
\end{equation*}
\begin{equation}
\left. {{\alpha _{j'i'}} \in \{ 1,...,n\} ,{\gamma _{{{q'}}i''}} \in \{ 1,...,sn\} ,{\beta _{{{q''}}i'''}} \in \{ 1,...,\upsilon n\} \begin{array}{*{20}{c}}
{}\\
{}
\end{array}} \right\},
\label{beam2}
\end{equation}
where $\tilde {\cal S}_1^{[i]}=\bar {\cal S}_1^{[i]}$ in (\ref{set1}), and we have:
\begin{equation}
\tilde {\cal S}_2^{[i]} = \left\{ {(i'',q')\left| {i'' \in \{ 1,...,K\} ,i'' \notin {{\cal B}_{q'}},q' \in \{ 1,...,Q\} } \right.} \right\},
\label{set5}
\end{equation}
\begin{equation}
\tilde {\cal S}_3^{[i]} = \left\{ {(i''',q'')\left| {i''' \in {{\cal B}_{q''}},q'' \in \{ 1,...,Q\} } \right.} \right\},
\label{set6}
\end{equation}
and ${{\bold{T}^{[{{q''}}i''']}}}$s  are $T\times T$ diagonal random matrices for each $i'''$ and $q''$, where each of  diagonal elements for each matrix is drawn independently from  a continuous cumulative probability distribution.

\vspace{10pt}

\textit{2) Beamforming matrix design for $i\in \{W+1,...,K\}$}:

We consider the beamforming matrix  ${{\bar{\bold{V}}}^{[i]}}$ as the following:
\begin{equation}
{\bar {\bf{V}} ^{[i]}} = \left\{ {\left[ {\prod\limits_{(i',j')\in\bar {\cal S}_1^{[i]}} {{{\left( {{{\bf{H}}^{[j'i']}}} \right)}^{{\alpha _{j'i'}}}}} } \right]\left[ {\prod\limits_{(i'',q')\in\bar {\cal S}_2^{[i]}} {{{\left( {{{\bf{H}}_{\rm T-IR}^{[{{q'}}i'']}}} \right)}^{{\gamma _{{{q'}}i''}}}}} } \right]{\bf w}:{\alpha _{j'i'}} \in \{ 1,...,n\} ,{\gamma _{{{q'}}i''}} \in \{ 1,...,tn\} } \right\},
\label{beam3}
\end{equation}
where $\bar {\cal S}_1^{[i]}$ and $\bar {\cal S}_2^{[i]}$ are given by (\ref{set1}) and (\ref{set2}), respectively.
 $t$ is a parameter for controlling the dimension of ${{\bar{\bold{V}}}^{[i]}}$, i.e., $d({{\bar{\bold{V}}}^{[i]}})$.

We note that  each value of parameters $s,\upsilon$ and $t$ can be approximated by rational numbers with arbitrarily small error, and by choosing a sufficiently large $n$, parameters $sn,\upsilon n$ and $tn$ will be integers and our proposed scheme will be realizable.

\vspace{10pt}

\textbf{Step 5: Validity of interference alignment conditions and decodability of message symbols}

Now, we analyze the spaces of messages and interference.

\textit{1) Validity of interference alignment conditions at clean receivers $j\in \{1,...,W\}$:}

For the clean receivers $j\in \{1,...,W\}$, we have the following lemma:

\begin{lemma}
For the clean receivers $j\in \{1,...,W\}$, consider $\bar{\cal{C}}_j$ as the message subspace corresponding to the symbol stream ${\bar{\bf x}}^{[j]}$,  $\tilde{\cal{C}}_j$ as the message subspace corresponding to the symbol stream ${\tilde{\bf x}}^{[j]}$, and  $\bar{\cal{A}}_j$ as the interference subspace induced by the symbol stream ${\bar{\bf x}}^{[j']},j'\ne j$. Then, $\bar{\cal{C}}_j,$ $\tilde{\cal{C}}_j$, and  $\bar{\cal{A}}_j$ are full rank and linearly independent, i.e., all base vectors of these subspaces are linearly independent. Thus, the message streams ${\bar{\bf x}}^{[j]}$ and ${\tilde{\bf x}}^{[j]}$ are decodable by zero forcing. In addition, we have:
\begin{equation}
D_N(\bar {\cal C}_j)=\Gamma,
\label{Ndim1}
\end{equation}
\begin{equation}
D_N(\tilde {\cal C}_j)=\chi,
\end{equation}
\begin{equation}
D_N({\bar{\cal A}}_j)=\max\{\Gamma,\zeta\},
\end{equation}
where 
\begin{equation*}
\Gamma  = {s^{QK}},\quad \chi  = {s^{QK-W}}{\upsilon ^W}, \quad \zeta  = {t^{QK}}.
\end{equation*}

\label{lemma_1}
\end{lemma}

\begin{IEEEproof}
The proof is provided in Appendix \ref{appendix1}.

\end{IEEEproof}


\textit{2) Validity of interference alignment conditions at dirty receivers $j\in \{W+1,...,K\}$:}

For the dirty receivers $j\in \{W+1,...,K\}$, we have the following lemma:

\begin{lemma}
For the dirty receivers $j\in \{W+1,...,K\}$, consider $\bar{\cal{C}}_j$ as the message subspace corresponding to the symbol stream ${\bar{\bf x}}^{[j]}$,  $\tilde{\cal{A}}_j$ as the interference subspace corresponding to the symbol stream ${\tilde{\bf x}}^{[j']},j'\ne j$, and  $\bar{\cal{A}}_j$ as the interference subspace induced by the symbol streams ${\bar{\bf x}}^{[j']},j'\ne j$. Then, $\bar{\cal{C}}_j, \tilde{\cal{A}}_j$, and  $\bar{\cal{A}}_j$ are full rank and linearly independent, i.e., all base vectors of these subspaces are linearly independent. Thus, the message stream ${\bar{\bf x}}^{[j]}$ is decodable by zero forcing. In addition, we have:
\begin{equation}
D_N ({\bar {\cal C}_j})=\zeta ,
\end{equation}
\begin{equation}
D_N({\bar {\cal A}_j})=\max\{\Gamma,\zeta\},
\end{equation}
\begin{equation}
D_N({\tilde {\cal A}_j})=\chi.
\end{equation}

\label{lemma_2}
\end{lemma}

\begin{IEEEproof}
The proof is provided in Appendix \ref{appendix2}.
\end{IEEEproof}


\textit{3) Validity of interference alignment conditions at the MIMO C-IR $q$-th receive antenna $q\in \{1,...,Q\}$:}

For the $q$-th receive antenna of the MIMO C-IR $q\in \{1,...,Q\}$, we have the following lemma:

\begin{lemma}
For the $q$-th receive antenna of the MIMO C-IR $q\in \{1,...,Q\}$, consider $\tilde {\cal{C}}_{i,r_q}$ as the message subspace corresponding to the symbol streams ${\tilde{\bf x}}^{[i]},i\in {\cal B}_q$,  $\tilde{\cal{A}}_{r_q}$ as the interference subspace corresponding to the symbol streams ${\tilde{\bf x}}^{[j]},j\ne {\cal B}_q$, and  $\bar{\cal{A}}_{r_q}$ as the interference subspace induced by the symbol streams ${\bar{\bf x}}^{[j]},\forall j$. Then, $\tilde {\cal{C}}_{i,r_q},i\in {\cal B}_q$  , $\bar{\cal{A}}_{r_q}$, and $\tilde{\cal{A}}_{r_q}$ are full rank and linearly independent,  i.e., all base vectors of these subspaces are linearly independent. Thus, the message stream ${\tilde{\bf x}}^{[i]},i\in {\cal B}_q$ can be demultiplexed by zero forcing. In addition, we have:
\begin{equation}
D_N(\tilde {\cal{C}}_{i,r_q})=\chi,
\end{equation}
\begin{equation}
\sum\limits_{i \in {{\cal B}_q}} {D_ N(\tilde {\cal{C}}_{i,r_q})}= \left| {{{\cal B}_q}} \right|\chi  ,
\end{equation}
\begin{equation}
D_N(\bar {\cal{A}}_{r_q})=\max\{\Gamma,\zeta\},
\end{equation}
\begin{equation}
D_N(\tilde {\cal{A}}_{r_q})=\chi.
\label{Ndim2}
\end{equation}

\label{lemma_3}
\end{lemma}

\begin{IEEEproof}
The proof is provided in Appendix \ref{appendix3}.
\end{IEEEproof}

Now, we can calculate the dimension of the whole signal space at each receiver. We define $d_{t,j}$ as the total dimension at the $j$-th receiver and $d_{t,r_q}$ as the total dimension at the $q$-th receive antenna of the MIMO C-IR, thus, we have:
\begin{equation}
d_{t,j}=d(\bar{\cal{C}}_j)+d(\tilde{\cal{C}}_j)+d(\bar{\cal{A}}_j),\forall j \in \{1,...,W\},
\label{d_{t,j}}
\end{equation}
\begin{equation}
d_{t,j}=d(\bar{\cal{C}}_j)+d(\bar{\cal{A}}_j)+d(\tilde{\cal{A}}_j),\forall j \in \{W+1,...,K\},
\end{equation}
\begin{equation}
d_{t,r_q}=\sum\limits_{i \in {{\cal B}_q}} {d({{\tilde {\cal C} }_{i,{r_q}}})} +d(\bar{\cal{A}}_{r_q})+d(\tilde{\cal{A}}_{r_q}),\forall q \in \{1,...,Q\},
\label{d_{t,r_q}}
\end{equation}
where the dimension of the message and interference subspaces are derived in (\ref{dim_mess_clean1})-(\ref{dim_inter_clean}), (\ref{d_2_1})-(\ref{d_2_3}), and (\ref{d_3_1})-(\ref{d_3_3}) in Appendices \ref{appendix1}-\ref{appendix3}.
Similarly, define $D_{N,t,j}$ as the total normalized asymptotic dimension at the $j$-th receiver and $D_{N,t,r_q}$ as the total normalized asymptotic dimension at the  $q$-th receive antenna of the MIMO C-IR, thus, from (\ref{Ndim1})-(\ref{Ndim2}),  we have:
\begin{equation}
D_{N,t,j}=D_N(\bar{\cal{C}}_j)+D_N(\tilde{\cal{C}}_j)+D_N(\bar{\cal{A}}_j)=\Gamma +\chi +\max\{\Gamma,\zeta\},\forall j \in \{1,...,W\},
\end{equation}
\begin{equation}
D_{N,t,j}=D_N(\bar{\cal{C}}_j)+D_N(\bar{\cal{A}}_j)+D_N(\tilde{\cal{A}}_j)=\zeta +\chi +\max\{\Gamma,\zeta\},\forall j \in \{W+1,...,K\},
\end{equation}
\begin{equation}
D_{N,t,r_q}=\sum\limits_{i \in {{\cal B}_q}} {D_N({{\tilde {\cal C} }_{i,{r_q}}})} +D_N(\bar{\cal{A}}_{r_q})+D_N(\tilde{\cal{A}}_{r_q})=\left| {{{\cal B}_q}} \right| \chi +\chi+\max\{\Gamma,\zeta\},\forall q \in \{1,...,Q\}.
\label{D_{N,t,r_q}}
\end{equation}
Now, we determine the minimum value for the parameter $T$ (for which the interference alignment equations are satisfied) as follows:
\begin{equation}
T = \max \left\{ {\mathop {\max }\limits_{j \in \{ 1,...,K\} } \{ {d_{t,j}}\} ,\mathop {\max }\limits_{q \in \{ 1,...,Q\} } \{ {d_{t,{r_q}}}\} } \right\},
\label{T_min}
\end{equation}
and from (\ref{d_{t,j}})-(\ref{T_min}), we have
\begin{equation}
\mathop {\lim }\limits_{n \to \infty }\frac {T}{n^{K^2-K+QK}} = \chi  + \max \left\{ {\Gamma ,\zeta } \right\} + \max \left\{ {\mathop {\max }\limits_{q \in \{ 1,...,Q\} } \left| {{{\cal B}_q}} \right|\chi ,\zeta ,\Gamma } \right\}.
\end{equation}
On the other hand, we have:
\begin{equation*}
\mathop {\max }\limits_{q \in \{ 1,...,Q\} } \left| {{{\cal B}_q}} \right| = \left\lceil {\frac{W}{Q}} \right\rceil ,
\end{equation*}
so, we conclude:
\begin{equation}
\mathop {\lim }\limits_{n \to \infty } \frac{T}{{{n^{{K^2} - K + QK}}}} = \chi  + \max \left\{ {\Gamma ,\zeta } \right\} + \max \left\{ {\left\lceil {\frac{W}{Q}} \right\rceil \chi ,\zeta ,\Gamma } \right\}.
\label{total l}
\end{equation}

Up to now, we have considered any arbitrary real values for each parameter $\Gamma,\chi$ and $\zeta$. Now, we make two additional assumptions on these parameters, which give us an achievable DoF. First, we set the normalized asymptotic dimension of the space at the clean receivers equal to that of the dirty receivers. Hence:
\begin{equation}
\Gamma = \zeta.
\label{assumption1}
\end{equation} 
Second, we set the maximum normalized asymptotic dimension of the space at each MIMO C-IR receive antenna to be less than or equal to that of the dirty receivers. Therefore, we have:
\begin{equation}
\zeta  \ge \left\lceil {\frac{W}{Q}} \right\rceil \chi.
\label{assumption2}
\end{equation}
Having (\ref{assumption1}) and (\ref{assumption2}), (\ref{total l}) will have the following form
\begin{equation}
\mathop {\lim }\limits_{n \to \infty } \frac{T}{{{n^{{K^2} - K + QK}}}} = \chi +2\Gamma.
\end{equation}

\vspace{10pt}

\textbf{Step 6: DoF analysis}

Now, we characterize the total DoF. As stated before, we have $W$ clean receivers each with normalized message dimension equal to $\Gamma+\chi$ and $K-W$ dirty receivers each with normalized message dimension equal to $\zeta$ (note that we set $\zeta=\Gamma$) and the total normalized transmission length is equal to $\chi +2\Gamma$, so the total DoF has the following form:
\begin{equation}
{\rm DoF} = \mathop {\max }\limits_{\chi  \ge 0,\Gamma  \ge \left\lceil {\frac{W}{Q}} \right\rceil \chi } \frac{{W(\chi  + \Gamma ) + (K - W)\Gamma }}{{\chi  + 2\Gamma }},
\label{optimization}
\end{equation}
and by assuming $\Gamma=\beta \chi$,  we have:
\begin{align}
{\rm DoF}& = \mathop {\max }\limits_{\beta  \ge \left\lceil {\frac{W}{Q}} \right\rceil } \frac{{W(1 + \beta ) + (K - W)\beta }}{{1 + 2\beta }}\\
 &= \frac{K}{2} + \mathop {\max }\limits_{\beta  \ge \left\lceil {\frac{W}{Q}} \right\rceil } K\frac{{\frac{W}{K} - \frac{1}{2}}}{{1 + 2\beta }} = \frac{K}{2} + \max \left\{ {K\frac{{\frac{W}{K} - \frac{1}{2}}}{{1 + 2\left\lceil {\frac{W}{Q}} \right\rceil }},0} \right\}.
\label{main result1}
\end{align}
We remark that if $\frac{W}{K} > \frac{1}{2}$, we set $\beta=\left\lceil {\frac{W}{Q}} \right\rceil$ and if $\frac{W}{K}<\frac{1}{2}$, we tend $\beta$ to $\infty$. This completes the proof of the achievability of the first term of (\ref{DoF1}). The proof of second term, i.e., $\min\{Q,W\}$ is provided  in Appendix \ref{appendix0}. 
\end{IEEEproof}

\textit{Remark 1}: It is known that the DoF is an appropriate performance metric, which provides a capacity approximation, accurate within $o(\log(\rho))$\cite{Cadambe1}. Therefore,
Theorem \ref{theorem1} indicates that the approximate sum capacity of a frequency-selective $K$-user interference channel in the presence of MIMO C-IR is lower bounded by $\left( {\max \left\{ {\frac{K}{2} + \max \left\{ {0,K\frac{{\frac{W}{K} - \frac{1}{2}}}{{1 + 2\left\lceil {\frac{W}{Q}} \right\rceil }}} \right\},\min \left\{ {Q,W} \right\}} \right\} - \epsilon } \right)\log (1 + \rho ) + o(\log (\rho )),\forall \epsilon  > 0$.
Now, we prove an improved achievable DoF for a special case of  $W$ and $Q$.    

\begin{theorem}
Assume $W=QZ+P,P=1$. Then, the achievable DoF (\ref{DoF1}) can be improved as follows:
\begin{equation}
{\rm DoF}= \max \left\{ {\frac{K}{2} + \max \left\{ {0,K\frac{{\frac{W}{K} - \frac{1}{2}}}{{1 + 2\left\lfloor {\frac{W}{Q}} \right\rfloor }}} \right\},\min \left\{ {Q,W} \right\}} \right\}.
\label{DoF2}
\end{equation}
\label{theorem2}

\end{theorem}

\begin{IEEEproof}
The proof is provided in Appendix \ref{appendix5}.
\end{IEEEproof}

\textit{Remark 2}: Theorem \ref{theorem2} shows that the approximate sum capacity of a frequency-selective $K$-user interference channel with a MIMO C-IR is lower bounded by $\left( {\max \left\{ {\frac{K}{2} + \max \left\{ {0,K\frac{{\frac{W}{K} - \frac{1}{2}}}{{1 + 2\left\lfloor {\frac{W}{Q}} \right\rfloor }}} \right\},\min \left\{ {Q,W} \right\}} \right\} - \epsilon } \right)$ $\log (1 + \rho ) + o(\log (\rho )),\forall \epsilon  > 0$, where $P=1$ (we have $W=QZ+P,0\le P<Q$). From (\ref{DoF1}) and (\ref{DoF2}), we note that this lower bound is tighter than the previous bound.

\textit{Remark 3}: As expected, if we set $Q=W=K$, the maximum $K$ DoF, which is the DoF at the absence of the interference, is achievable for the MIMO C-IR.

\textit{Remark 4}: It has been shown in \cite{Cadambe3} that an ordinary relay cannot increase the DoF of the $K$-user interference channel. The main difference here is that the instantaneity of the relay can significantly improve the DoF.

\section{Frequency-Selective $K$-User Interference Channel in the Presence of NC-IR}
\label{sb}

In this section, we provide an achievability theorem for the DoF of the frequency-selective $K$-user interference channel in the presence of an NC-IR as follows:
\begin{theorem}
Consider $U,p,e,e'\in \mathbb{W}$ such that
\begin{equation}
U=pe+e',0\le e' <p,\frac{K}{2}<U\le K.
\end{equation}
Then, with an NC-IR with $W=Q=pU$ antennas, the following DoF is achievable:
\begin{equation}
{\rm DoF} = \frac{K}{2} + \max \left\{ {K\frac{{\frac{U}{K} - \frac{1}{2}}}{{1 + 2\left\lceil {\frac{U}{p}} \right\rceil }},0} \right\}.
\label{NC-IR-achievable DoF}
\end{equation}
\label{theorem3}
\end{theorem}

\begin{IEEEproof}
The proof is provided in Appendix \ref{appendix6}.
\end{IEEEproof}

\textit{Remark 5}: 
Theorem \ref{theorem3} indicates that the approximate sum capacity of a frequency-selective $K$-user interference channel in the presence of NC-IR is lower bounded by $\left( {\frac{K}{2} + \max \left\{ {K\frac{{\frac{U}{K} - \frac{1}{2}}}{{1 + 2\left\lceil {\frac{U}{p}} \right\rceil }},0} \right\} - \epsilon } \right)\log (1 + \rho ) + o(\log (\rho )),\forall \epsilon  > 0$.

\textit{Remark 6}: The active intelligent reflecting surface, can be modeled as a special case of NC-IR \cite{me}. It has been proven in \cite{me} that for an active intelligent reconfigurable surface with $Q=U(K-1)+U(K-U)$ antennas, the following DoF is achievable:
\begin{equation}
{\rm DoF}=\frac{K+U}{2},0\le U\le K.
\label{active IRS}
\end{equation}
 Therefore, we can see that for $0<Q<2(K-1)$, the achievable DoF (\ref{NC-IR-achievable DoF}) is dominant, and for $Q\ge2(K-1)$, the maximum of (\ref{NC-IR-achievable DoF}) and (\ref{active IRS}) forms the maximum achievable DoF for the NC-IR.

\textit{Remark 7}: From Theorem \ref{theorem1}, we conclude that the maximum $K$ DoF can be achieved by $Q=W=K$ antennas for a MIMO C-IR, but the number of antennas for achieving the maximum $K$ DoF by an NC-IR is $Q=K(K-1)$, which grows quadratically and shows the loss of performance.

\section{Numerical Results}
\label{section4}

\begin{figure}
\centering
\includegraphics[width=12cm]{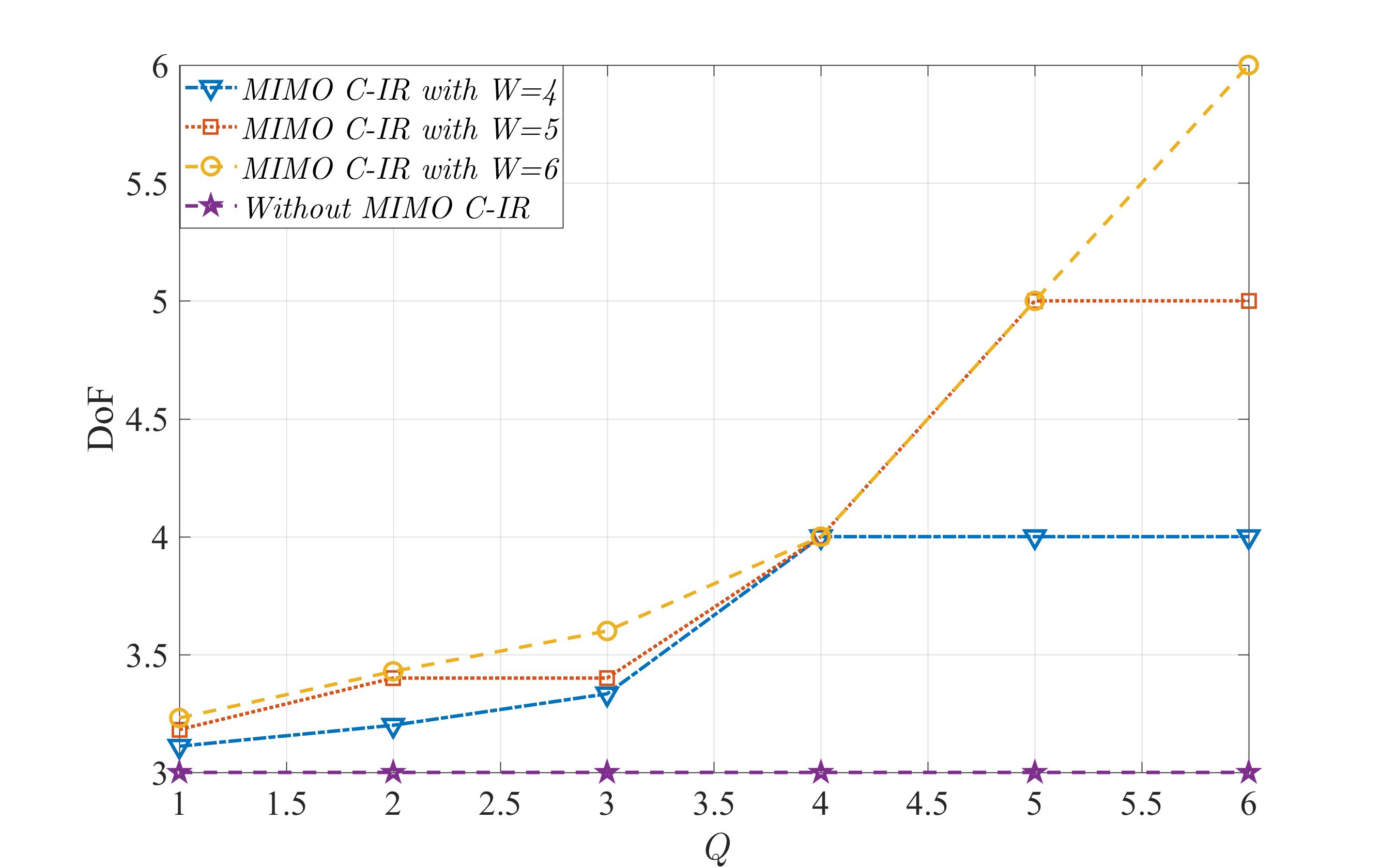}
\caption{Comparison of the achievable DoF for $6$-user interference channel in the presence of MIMO C-IR with the case without MIMO C-IR.}
 \label{fig0}
\end{figure}

In this section, we numerically evaluate the achievable DoFs provided in previous sections, through some examples.
In Fig. \ref{fig0}, we compare the achievable DoF for the $6$-user interference channel in the presence of MIMO C-IR for different values of $Q$ and $W$ and the case without MIMO C-IR. We see that the achievable DoF can only approach maximum value ($K=6$) when $W=K=6$. Also, it is observed  that the maximum achieved DoF is equal to $W$, when $W\ge4$. Moreover, the maximum $K$ DoFs can be achieved, when $Q=W$.

\begin{figure}
\centering
\includegraphics[width=12cm]{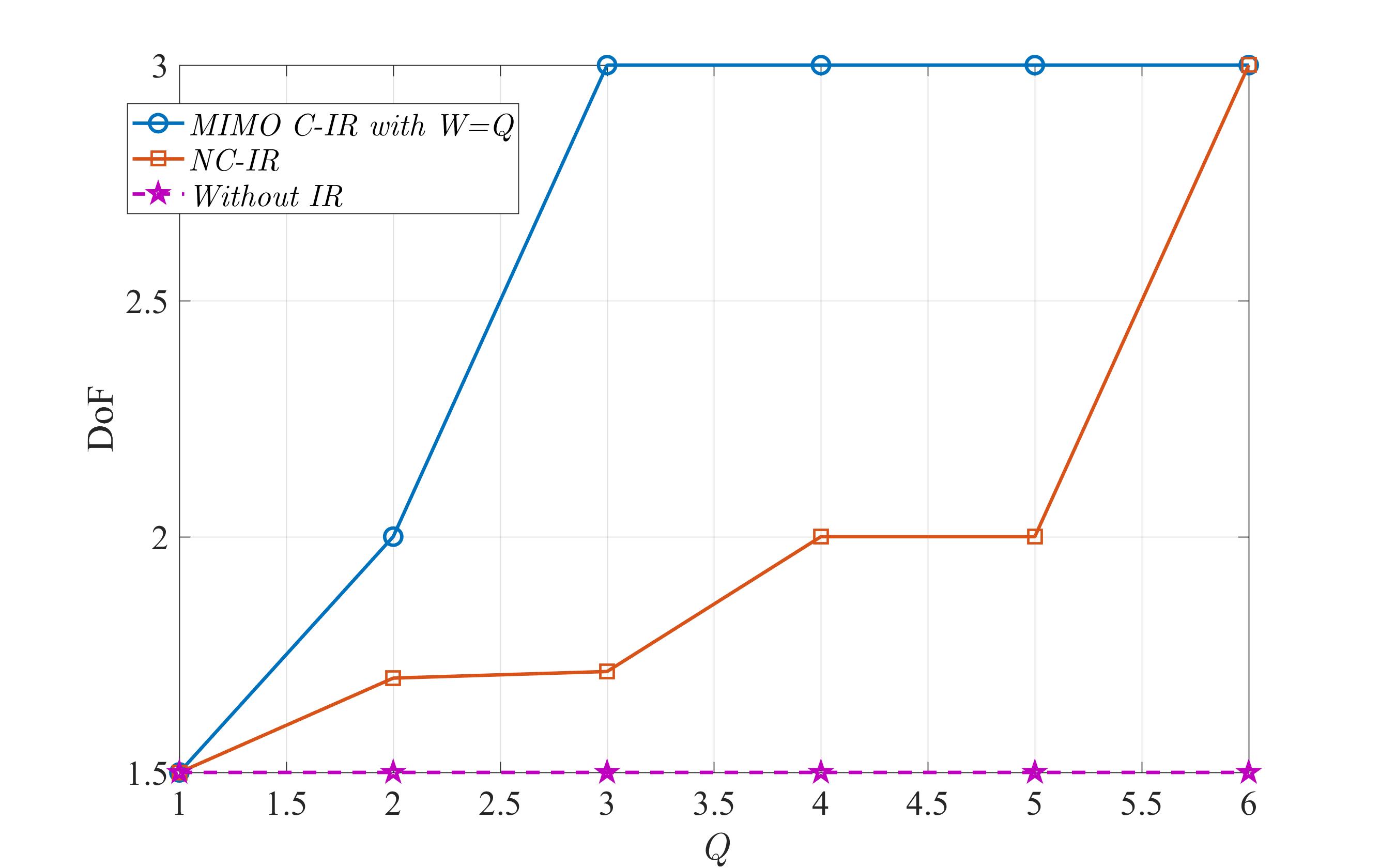}
\caption{Comparison of the achievable DoF for $3$-user interference channel in the presence of MIMO C-IR (with $W=Q$), NC-IR with the case without MIMO C-IR.}
 \label{fig1}
\end{figure}

\begin{figure}
\centering
\includegraphics[width=12cm]{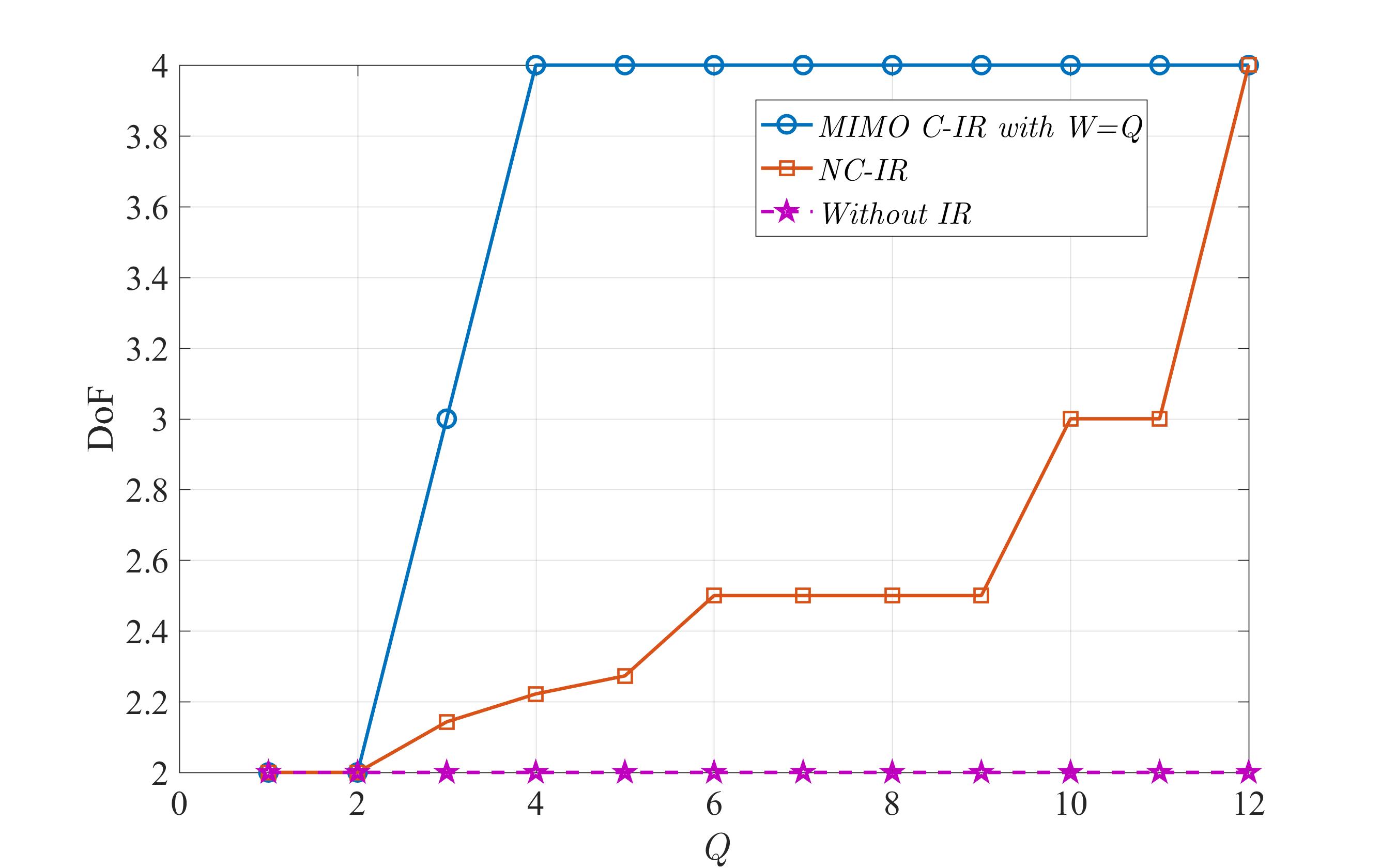}
\caption{Comparison of the achievable DoF for $4$-user interference channel in the presence of MIMO C-IR (with $W=Q$),  NC-IR with the case without MIMO C-IR.}
 \label{fig2}
\end{figure}

In Fig.s \ref{fig1} and \ref{fig2}, we compare the achievable DoF for $3$ and $4$-user interference channels in the presence of MIMO C-IR, NC-IR, and the case without IR. We note that to have a fair comparison, we assume the same number of receive and transmit antennas for the MIMO C-IR ($W=Q$) as it is for the NC-IR. These figures show that the maximum $K$ DoF can be achieved by employing enough number of antennas for both MIMO C-IR and NC-IR. We see that the achievable DoF is considerably decreased for the NC-IR, and this reduction is due to the lack of coordination between the antennas in the NC-IR.
Also, these figures show that the required number of antennas for an NC-IR to achieve the maximum $K$ DoF is quadratically larger than the required number of antennas for a MIMO C-IR, which shows the performance loss for an NC-IR due to lack of coordination between NC-IR antennas.


\section{Conclusion}
\label{section5}

In this paper, we studied achievable DoF for IR-assisted frequency-selective $K$-user interference channel and proposed novel interference alignment-based coding schemes. For a MIMO C-IR, whose antennas can have coordination with each other, and for NC-IR (an IR with no coordination between antennas), we derived achievable DoFs and observed the performance loss for the NC-IR compared with the MIMO C-IR. Also, we showed that by considering the number of antennas more than a finite value, the maximum $K$ DoF is achievable for both MIMO C-IR and NC-IR. The direction of our future works contains the following aspects:
1) finding tight bounds for the DoF of time-selective $K$-user interference channel in the presence of IR, 
2) extending our proposed coding scheme for more general wireless channels, e.g. $X$-network, and
3) extending our coding scheme to the scenario with imperfect CSI.

\begin{appendices}
\section{Proof of the achievability of the  second term $\min\{Q,W\}$ in  Theorem \ref{theorem1}}
\label{appendix0}
In this scheme, we only use one frequency slot $\omega_1$. We set $L=\min\{W,Q\}$. We assume that only transmitters  $i\in \{1,...,L\}$ send their messages to the receivers $j\in \{1,...,L\}$ via symbols $X^{[i]}(\omega_1),i\in\{1,...,L\}$ and other transmitters are silent ($X^{[i]}(\omega_1)=0,\forall i \in \{L+1,...,K\}$). From (\ref{model relay}), the MIMO C-IR can demultiplex $X^{[i]}(\omega_1),\forall i \in \{1,...,L\}$ by $L$ linear equations in first $L$ receive antennas almost surely, because the matrix of the coefficients is in terms of independent random variables, thus, its determinant is a non-zero polynomial of independent random variables with a continuous cumulative probability distribution and by \cite[Lemma 1]{me}, it is non-zero with probability $1$. Then, the MIMO C-IR designs
its transmitted signal to remove the interference in each receiver $j\in\{1,...,L\}$ by solving the following linear equations:
\begin{equation}
- \sum\limits_{i \in \{ 1,...,L\} ,i \ne j} {{H^{[ji]}}({\omega _1})({X^{[i]}}({\omega _1})+{{\tilde Z}^{[i]}}({\omega _t}))}  = \sum\limits_{u = 1}^L {{H_{\rm IR-R}^{[ju]}}{X_{\rm IR}^{[{u}]}}} ,\forall j \in \left\{ {1,...,L} \right\},
\label{schem2}
\end{equation}
\begin{equation}
{{\tilde {\tilde Z}}^{[j]}}({\omega _t}) =  - \sum\limits_{i \in \{ 1,...,L\} ,i \ne j} {{H^{[ji]}}({\omega _1}){{\tilde Z}^{[i]}}({\omega _t})} .
\end{equation}
where ${{\tilde Z}^{[i]}}({\omega _t})$ is the detection noise for symbol ${{X}^{[i]}}({\omega _t})$ at MIMO C-IR. Note that by this procedure, the interference cancellation is done, but we have an additional noise ${{\tilde {\tilde Z}}^{[j]}}({\omega _t}) $, which is negligible in high signal to noise ratio (SNR) regime.
Therefore, $L$ symbols can be transmitted in one frequency slot and the total $L$ DoFs are achievable. Thus, the second term in (\ref{DoF1}) is achievable, which completes the proof.

\section{Proof of Lemma \ref{lemma_1}}
\label{appendix1}
Using (\ref{beam1}) and (\ref{beam2}), we characterize the message  subspaces $\bar{\cal{C}}_j$ and $\tilde{\cal{C}}_j$ as follows:
\begin{fleqn}
\begin{equation*}
{\bar {\cal C}_j} = \textrm{span}\left( {{{\bold H}^{[jj]}}{{\bar {\bold V}}^{[j]}}} \right)=
\end{equation*}
  \end{fleqn}
\begin{equation}
\textrm{span} \left\{ {\left[ {\prod\limits_{(i',j') \in \bar {\cal S}_{1,j}^{\cal C}} {{{\left( {{{\bold H}^{[j'i']}}} \right)}^{{\alpha _{j'i'}}}}} } \right]\left[ {\prod\limits_{(i'',q') \in \bar {\cal S}_{2,j}^{\cal C}} {{{\left( {{{\bold H}_{\rm T-IR}^{[{{q'}}i'']}}} \right)}^{{\gamma _{{{q'}}i''}}}}} } \right]{\bf w}:{\alpha _{j'i'}} \in \bar {\cal S}_{{j'i'j}}^{\alpha ,{\cal C}},{\gamma _{{{q'}}i''}} \in \bar {\cal S}_{{r_{q'}}i''j}^{\gamma ,{\cal C}}} \right\},
\label{c_bar_clean}
\end{equation}
where $\bar {\cal S}_{2,j}^{\cal{C}}=\bar {\cal S}_2^{[i]}$ in (\ref{set2}), and sets $\bar {\cal S}_{1,j}^{\cal{C}}, \bar {\cal S}_{{j'i'j}}^{\alpha,{\cal{C}}} $, and $\bar {\cal S}_{{r_{q'}}i''j}^{\gamma,{\cal{C}}}  $ are defined as follows: 
\begin{equation}
\bar {\cal S}_{1,j}^{\cal C} = \left\{ {(i',j')\left| {i',j' \in \{ 1,...,K\} } \right.} \right\},
\label{set4}
\end{equation}
\begin{equation}
\bar {\cal S}_{j'i'j}^{\alpha ,{\cal C}} =  {\begin{cases}
{\{ 1,...,n\} ,i' \ne j'}\\
{\{ 0\} ,j' \ne j,i' = j'}\\
{\{ 1\} ,j' = j,i' = j}
\end{cases}} ,
\label{set7}
\end{equation}
\begin{equation}
\bar {\cal S}_{{r_{q'}}i''j}^{\gamma ,{\cal C}} = \left\{ {1,...,sn} \right\}.
\end{equation}
\begin{fleqn}
\begin{equation*}
{\tilde {\cal C}_j} = \textrm{span}\left( {{{\tilde{\bold H}}^{[jj]}}{{\tilde {\bold V}}^{[j]}}} \right)=
\end{equation*}
\end{fleqn}
\begin{equation*}
\textrm{span} \left\{ {\left[ {\prod\limits_{(i',j') \in \tilde {\cal S}_{1,j}^{\cal C}} {{{\left( {{{\tilde{\bold H}}^{[j'i']}}} \right)}^{{\alpha _{j'i'}}}}} } \right]\left[ {\prod\limits_{(i'',q') \in \tilde {\cal S}_{2,j}^{\cal C}} {{{\left( {{{\bold H}_{\rm T-IR}^{[{{q'}}i'']}}} \right)}^{{\gamma _{{{q'}}i''}}}}} } \right]\left[ {\prod\limits_{(i''',q'') \in \tilde {\cal S}_{3,j}^{\cal C}} {{{\left( {{{\bold T}^{[{{q''}}i''']}}} \right)}^{{\beta _{{{q''}}i'''}}}}} } \right]} \right.{\bf w}:
\end{equation*}
\begin{equation}
\left. {{\alpha _{j'i'}} \in \tilde {\cal S}_{{j'i'j}}^{\alpha ,{\cal C}},{\gamma _{{{q'}}i''}} \in \tilde {\cal S}_{{r_{q'}}i''j}^{\gamma ,{\cal C}},{\beta _{{{q''}}i'''}} \in \tilde {\cal S}_{{r_{q''}}i'''j}^{\beta ,{\cal C}}\begin{array}{*{20}{c}}
{}\\
{}
\end{array}} \right\},
\label{c_tilde_clean}
\end{equation}
where $\tilde {\cal S}_{1,j}^{\cal C}=\bar {\cal S}_{1,j}^{\cal C}$ in (\ref{set4}), $\tilde {\cal S}_{2,j}^{\cal C}=\tilde {\cal S}_2^{[i]}$ in (\ref{set5}), and $\tilde {\cal S}_{3,j}^{\cal C}=\tilde {\cal S}_3^{[i]}$ in (\ref{set6}).  $\tilde {\cal S}_{{j'i'j}}^{\alpha ,{\cal C}}$, $\tilde {\cal S}_{{r_{q'}}i''j}^{\gamma ,{\cal C}}$, and $\tilde {\cal S}_{{r_{q''}}i'''j}^{\beta ,{\cal C}}$ are defined as follows:
\begin{equation}
\tilde {\cal S}_{j'i'j}^{\alpha ,{\cal C}} = {\begin{cases}
{\{ 1,...,n\} ,i' \ne j'}\\
{\{ 0\} ,j' \ne j,i' = j'}\\
{\{ 1\} ,j' = j,i' = j}
\end{cases}} ,
\end{equation}
\begin{equation}
\tilde {\cal S}_{{r_{q'}}i''j}^{\gamma ,{\cal C}} = \left\{ {1,...,sn} \right\},
\end{equation}
\begin{equation}
\tilde {\cal S}_{{r_{q''}}i'''j}^{\beta ,{\cal C}} = \left\{ {1,...,\upsilon n} \right\}.
\end{equation}

To satisfy the interference alignment equation (\ref{IA1}), the subspace $\bar{\cal A}_j$ must be chosen such that:
\begin{equation*}
\bigcup\limits_{i\in\{1,...,K\},i \ne j} {\left\{ {{\text{span}}\left( {{{\mathbf{H}}^{[ji]}}{{\bar {\mathbf{V}} }^{[i]}}} \right)} \right\}}  \subseteq {\bar {\cal A} _j}.
\end{equation*}
Therefore, we characterize ${{\bar {\cal{A}}}_j} $ as follows:
\begin{equation}
{{\bar {\cal{A}}}_j} =\textrm{span} \left\{ {\left[ {\prod\limits_{(i',j') \in \bar {\cal S}_{1,j}^{\cal{A}}} {{{\left( {{{\bold H}^{[j'i']}}} \right)}^{{\alpha _{j'i'}}}}} } \right]\left[ {\prod\limits_{(i'',q') \in \bar {\cal S}_{2,j}^{\cal{A}}} {{{\left( {{{\bold H}_{\rm T-IR}^{[{{q'}}i'']}}} \right)}^{{\gamma _{{{q'}}i''}}}}} } \right]{\bf w}:{\alpha _{j'i'}} \in \bar {\cal S}_{{j'i'j}}^{\alpha,{\cal{A}}} ,{\gamma _{{{q'}}i''}} \in \bar {\cal S}_{{r_{q'}}i''j}^{\gamma,{\cal{A}}} } \right\}
\label{a_bar_clean}
\end{equation}
where $\bar {\cal S}_{1,j}^{\cal{A}}=\bar {\cal S}_1^{[i]}$ in (\ref{set1}) and $\bar {\cal S}_{2,j}^{\cal{A}}=\bar {\cal S}_2^{[i]}$ in (\ref{set2}). $\bar {\cal S}_{{j'i'j}}^{\alpha,{\cal{A}}} $ and $\bar {\cal S}_{{r_{q'}}i''j}^{\gamma,{\cal{A}}}  $ are defined as follows: 
\begin{equation}
\bar {\cal S}_{{j'i'j}}^{\alpha,{\cal{A}}}  =  {\begin{cases}
{\{ 1,...,n\} ,j' \ne j}\\
{\{ 1,...,n + 1\} ,j' = j}
\end{cases}} ,
\end{equation}
\begin{equation}
\bar {\cal S}_{{r_{q'}}i''j}^{\gamma,{\cal{A}}}  = \left\{ {1,...,\max \{ sn,tn\} } \right\}.
\end{equation}
Note that to use the zero-forcing technique, the subspace of the interference must be a vector space, but the set of interference vectors, which is equal to $\bigcup\limits_{i\in\{1,...,K\},i \ne j} {\left\{ {{\text{span}}\left( {{{\mathbf{H}}^{[ji]}}{{\bar {\mathbf{V}} }^{[i]}}} \right)} \right\}}$, is not a vector space, thus, we choose the subspace of interference as (\ref{a_bar_clean}), which is easier to work with and includes $\bigcup\limits_{i\in\{1,...,K\},i \ne j} {\left\{ {{\text{span}}\left( {{{\mathbf{H}}^{[ji]}}{{\bar {\mathbf{V}} }^{[i]}}} \right)} \right\}}$.

After that, we analyze the dimension and the normalized asymptotic dimension of the messages and interference subspaces. First, we assume that the parameter $T$ (the number of frequency slots) is sufficiently large and at the end of step 5 of the proof, we will choose the minimum value for $T$ such that all message streams can be decodable and all interference alignment equations are satisfied. By the nature of $\bar {\cal {A}}_j$ in (\ref{a_bar_clean}), $\bar {\cal C}_j$ in (\ref{c_bar_clean}) and $\tilde {\cal C}_j$ in (\ref{c_tilde_clean}), we can see from the statement of the \cite[Lemma~2]{me}, if we choose the variables $x_k$ as $H^{[ji]}(\omega_t),H_{\rm T-IR}^{[qi']}(\omega_t),i,i',j\in \{1,...,K\},q\in\{1,...,Q\}$, $y_k$ as $H_{\rm IR-R}^{[ju]}(\omega_t),j\in \{W+1,...,K\},u\in \{1,...,W\}$ and $z_k$ as $H_{\rm IR-R}^{[ju]}(\omega_t),j\in \{1,...,W\},u\in \{1,...,W\}$, then by \cite[Lemmas 1-3]{me},  subspaces $\bar {\cal {A}}_j$, $\bar {\cal C}_j$ and $\tilde {\cal C}_j$ are almost surely full rank  and linearly independent (all base vectors of these subspaces are linearly independent). In fact, if we take the constructing base vectors of $\bar {\cal {A}}_j$, $\bar {\cal C}_j$ and $\tilde {\cal C}_j$ and construct a square matrix by choosing some rows of it, we can see by \cite[Lemmas 2-3]{me} that the determinant of this square matrix will be a non-zero polynomial and by \cite[Lemma 1]{me}, it will  be non-zero with probability equal to one, thus, all message streams are decodable at the clean receivers (by zero forcing).
For more clarity, we review \cite[Lemmas 1-3]{me} as follows:

\cite[Lemma 1]{me}:
Consider $k$ independent random variables $X_1,...,X_k$, each constructed from a continuous cumulative probability distribution. The probability of the event that a nonzero polynomial $P_k(X_1,...,X_k)$ constructed from $X_1,...,X_k$  with finite degree assumes the value zero is zero, i.e., $\Pr\{P_k(X_1,...,X_k)=0\}=0$.

\cite[Lemma 2]{me}:
Consider three sets of variables $\{{x_i},i \in {{\cal A}_x},\left| {{{\cal A}_x}} \right| < \infty \}$, $\{{y_i},i \in {{\cal A}_y},\left| {{{\cal A}_y}} \right| < \infty\}$, and $\{{z_i},i \in {{\cal A}_z},\left| {{{\cal A}_z}} \right| < \infty \}$. Consider  the following functions:
\begin{equation}
{f_j} = \prod\limits_{i = 1}^{\left| {{{\cal A}_x}} \right|} {{{\left( {{x_i} + \sum\limits_{i' \in {{\cal C}_j},i'' \in {{\cal D}_j}} {{x_{i'}}{y_{i''}}{P_1}^{[i'i''j]}({z_k}:k \in {{\cal A}_z}) + {y_{i''}}{P_2}^{[i'i''j]}({z_k}:k \in {{\cal A}_z})} } \right)}^{a_i^j}}},
\end{equation}
\begin{equation*}
(a_1^j,...,a_{\left| {{{\cal A}_x}} \right|}^j) \in {\mathbb{W}^{\left| {{{\cal A}_x}} \right|}},j \in \{ 1,...,J\},
\end{equation*}
where ${P_1^{[i'i''j]}}(\cdot)$ and ${P_2^{[i'i''j]}}(\cdot)$ are fractional polynomials and for $\forall j$, we have $\left| {{{\cal C}_j}} \right|,\left| {{{\cal D}_j}} \right| < \infty$. If for $\forall j,j'$ with $j\ne j'$, $(a_1^j,...,a_{\left| {{{\cal A}_x}} \right|}^j) \ne (a_1^{j'},...,a_{\left| {{{\cal A}_x}} \right|}^{j'})$, then the functions $f_j$ will be linearly independent.

\cite[Lemma 3]{me}:
Consider the set of nonzero linearly independent fractional polynomials $\{P^{[j]}(\cdot),j\in\{1,...,J\}\}$ and consider $J$ sets of variables ${{{\cal X}}}_j=\{x_i^j:i \in {\cal I},{\cal I}\subseteq \mathbb{N},|{\cal I}|<\infty\}$, $j \in \{1,...,J\}$. The determinant of the following matrix will be a nonzero fractional polynomial:
\begin{equation}
\bold{A}=\left[ {\begin{array}{*{20}{c}}
{{P^{[1]}}({{{\cal X}}_1})}&{{P^{[2]}}({{{{\cal X}}}_1})}& \cdots &{{P^{[J]}}({{{{\cal X}}}_1})}\\
{{P^{[1]}}({{{{\cal X}}}_2})}&{{P^{[2]}}({{{{\cal X}}}_2})}& \cdots &{{P^{[J]}}({{{{\cal X}}}_2})}\\
 \vdots & \vdots & \ddots & \vdots \\
{{P^{[1]}}({{{{\cal X}}}_J})}&{{P^{[2]}}({{{{\cal X}}}_J})}& \cdots &{{P^{[J]}}({{{{\cal X}}}_J})}
\end{array}} \right].
\end{equation}

\vspace{20pt}

 Now, we have to make sure that interference alignment equations (\ref{IA1}) and (\ref{IA2}) are satisfied by analyzing the dimension of message streams and interference. The dimension of the message subspaces $\bar {\cal C}_j$ and $\tilde {\cal C}_j$, which is equal to the number of its base vectors in (\ref{c_bar_clean}) and (\ref{c_tilde_clean}), can be characterized as follows:
\begin{equation}
d(\bar {\cal C}_j)= {n^{{K^2} - K}}{(sn)^{QK}},
\label{dim_mess_clean1}
\end{equation}
\begin{equation}
d(\tilde {\cal C}_j) ={n^{{K^2} - K}}{(sn)^\varphi }{(\upsilon n)^\theta },
\label{dim_mess_clean2}
\end{equation}
where
\begin{equation*}
\varphi  = \sum\limits_{q' = 1}^Q {(K - \left| {{{\cal B}_{q'}}} \right|)}  = KQ - \sum\limits_{q' = 1}^Q {\left| {{{\cal B}_{q'}}} \right|}  = KQ - W,
\end{equation*}
\begin{equation*}
\theta  = \sum\limits_{q' = 1}^Q {\left| {{{\cal B}_{q'}}} \right|}  = W.
\end{equation*}

The dimension of the interference subspace ${\bar{\cal A}}_j$, which is equal to the number of its base vectors in (\ref{a_bar_clean}) is:
\begin{equation}
d({\bar{\cal A}}_j)={n^{{K^2} - K - (K - 1)}}{(n + 1)^{K - 1}}{\left( {\max \{ sn,tn\} } \right)^{QK}}.
\label{dim_inter_clean}
\end{equation}
We can see  from (\ref{dim_mess_clean1})-(\ref{dim_inter_clean}) and (\ref{DN}) that $l=K^2-K+QK$.
We define the following parameters:
\begin{equation}
\Gamma  = {s^{QK}},
\end{equation}
\begin{equation}
\chi  = {s^{QK-W}}{\upsilon ^W},
\end{equation}
\begin{equation}
\zeta  = {t^{QK}}.
\label{asym dim3}
\end{equation}
From (\ref{dim_mess_clean1})-(\ref{asym dim3}) and (\ref{DN}), the normalized asymptotic dimension of the message and interference subspaces are:
\begin{equation}
D_N(\bar {\cal C}_j)=\Gamma,
\end{equation}
\begin{equation}
D_N(\tilde {\cal C}_j)=\chi,
\end{equation}
\begin{equation}
D_N({\bar{\cal A}}_j)=\max\{\Gamma,\zeta\}.
\end{equation}
The  interference alignment equations (\ref{IA1}) and (\ref{IA2}) are satisfied because we can  see  that the normalized asymptotic dimension of  the interference induced by ${\bar {\bold V}}^{[i]} {\bar {\bold x}}^{[i]},i\in \{1,..,W\},i \ne j$ is $\Gamma$ and  the normalized asymptotic dimension of  the interference induced by ${\bar {\bold V}}^{[i]} {\bar {\bold x}}^{[i]},i\in \{W+1,..,K\}$ is $\zeta$.
\section{Proof of Lemma \ref{lemma_2}}
\label{appendix2}
Using (\ref{beam3}), we characterize the message  subspace $\bar{\cal{C}}_j$ as follows:
\begin{fleqn}
\begin{equation*}
{\bar {\cal C}_j} = \textrm{span}\left( {{{\bold H}^{[jj]}}{{\bar {\bold V}}^{[j]}}} \right)=
\end{equation*}
\end{fleqn}
\begin{equation}
\textrm{span} \left\{ {\left[ {\prod\limits_{(i',j') \in \bar {\cal S}_{1,j}^{\cal C}} {{{\left( {{{\bold H}^{[j'i']}}} \right)}^{{\alpha _{j'i'}}}}} } \right]\left[ {\prod\limits_{(i'',q') \in \bar {\cal S}_{2,j}^{\cal C}} {{{\left( {{{\bold H}_{\rm T-IR}^{[{{q'}}i'']}}} \right)}^{{\gamma _{{{q'}}i''}}}}} } \right]{\bf w}:{\alpha _{j'i'}} \in \bar {\cal S}_{{j'i'j}}^{\alpha ,{\cal C}},{\gamma _{{{q'}}i''}} \in \bar {\cal S}_{{r_{q'}}i''j}^{\gamma ,{\cal C}}} \right\},
\label{c_bar_dirty}
\end{equation}
where $\bar {\cal S}_{1,j}^{\cal{C}}=\bar {\cal S}_{1,j}^{\cal C}$ in (\ref{set4}) and $\bar {\cal S}_{2,j}^{\cal{C}}=\bar {\cal S}_2^{[i]}$ in (\ref{set2}). $\bar {\cal S}_{{j'i'j}}^{\alpha,{\cal{C}}} $ and $\bar {\cal S}_{{r_{q'}}i''j}^{\gamma,{\cal{C}}}  $ are defined as follows: 
\begin{equation}
\bar {\cal S}_{j'i'j}^{\alpha ,{\cal C}} =  {\begin{cases}
{\{ 1,...,n\} ,i' \ne j'}\\
{\{ 0\} ,j' \ne j,i' = j'}\\
{\{ 1\} ,j' = j,i' = j}
\end{cases}},
\end{equation}
\begin{equation}
\bar {\cal S}_{{r_{q'}}i''j}^{\gamma ,{\cal C}} = \left\{ {1,...,tn} \right\}.
\end{equation}
To satisfy the interference alignment equation (\ref{IA3}), the subspace $\bar{\cal A}_j$ must be chosen such that:
\begin{equation*}
\bigcup\limits_{i\in\{1,...,K\},i \ne j} {\left\{ {{\text{span}}\left( {{{\mathbf{H}}^{[ji]}}{{\bar {\mathbf{V}} }^{[i]}}} \right)} \right\}}  \subseteq {\bar {\cal A} _j}.
\end{equation*}
Therefore, we characterize ${{\bar {\cal{A}}}_j} $ as follows:
\begin{equation}
{{\bar {\cal{A}}}_j} =\textrm{span} \left\{ {\left[ {\prod\limits_{(i',j') \in \bar {\cal S}_{1,j}^{\cal{A}}} {{{\left( {{{\bold H}^{[j'i']}}} \right)}^{{\alpha _{j'i'}}}}} } \right]\left[ {\prod\limits_{(i'',q') \in \bar {\cal S}_{2,j}^{\cal{A}}} {{{\left( {{{\bold H}_{\rm T-IR}^{[{{q'}}i'']}}} \right)}^{{\gamma _{{{q'}}i''}}}}} } \right]{\bf w}:{\alpha _{j'i'}} \in \bar {\cal S}_{{j'i'j}}^{\alpha,{\cal{A}}} ,{\gamma _{{{q'}}i''}} \in \bar {\cal S}_{{r_{q'}}i''j}^{\gamma,{\cal{A}}} } \right\}
\label{a_bar_dirty}
\end{equation}
where $\bar {\cal S}_{1,j}^{\cal{A}}=\bar {\cal S}_1^{[i]}$ in (\ref{set1}) and $\bar {\cal S}_{2,j}^{\cal{A}}=\bar {\cal S}_2^{[i]}$ in (\ref{set2}). $\bar {\cal S}_{{j'i'j}}^{\alpha,{\cal{A}}} $, and $\bar {\cal S}_{{r_{q'}}i''j}^{\gamma,{\cal{A}}}  $ are defined as follows: 
\begin{equation}
\bar {\cal S}_{{j'i'j}}^{\alpha,{\cal{A}}}  =  {\begin{cases}
{\{ 1,...,n\} ,j' \ne j}\\
{\{ 1,...,n + 1\} ,j' = j}
\end{cases}} ,
\end{equation}
\begin{equation}
\bar {\cal S}_{{r_{q'}}i''j}^{\gamma,{\cal{A}}}  = \left\{ {1,...,\max \{ sn,tn\} } \right\}.
\end{equation}
To satisfy the interference alignment equation (\ref{IA5}), the subspace $\tilde{\cal A}_j$ must be chosen such that:
\begin{equation*}
\bigcup\limits_{i\in\{1,...,W\}} {\left\{ {{\text{span}}\left( {{\tilde{\mathbf{H}}^{[ji]}}{{\tilde {\mathbf{V}} }^{[i]}}} \right)} \right\}}  \subseteq {\tilde {\cal A} _j}.
\end{equation*}
Therefore, we characterize subspace ${{\bar {\cal{A}}}_j} $ as follows:
\begin{equation*}
\tilde {\cal A}_j =\textrm{span} \left\{ {\left[ {\prod\limits_{(i',j')\in\tilde {\cal S}_{1,j}^{\cal A}} {{{\left( {{{{\tilde{\bold H}}}^{[j'i']}}} \right)}^{{\alpha _{j'i'}}}}} } \right]\left[ {\prod\limits_{(i'',q')\in\tilde {\cal S}_{2,j}^{\cal A}} {{{\left( {{{\bf{H}}_{\rm T-IR}^{[{{q'}}i'']}}} \right)}^{{\gamma _{{{q'}}i''}}}}} } \right]\left[ {\prod\limits_{(i''',q'')\in\tilde {\cal S}_{3,j}^{\cal A}} {{{\left( {{\bold{T}^{[{{q''}}i''']}}} \right)}^{{\beta _{{{q''}}i'''}}}}} } \right]{\bf w}:} \right.
\end{equation*}
\begin{equation}
\left. {{\alpha _{j'i'}} \in \tilde {\cal S}_{{j'i'j}}^{\alpha,{\cal{A}}} ,{\gamma _{{{q'}}i''}} \in \tilde {\cal S}_{{j'i'j}}^{\gamma,{\cal{A}}} ,{\beta _{{{q''}}i'''}} \in \tilde {\cal S}_{{j'i'j}}^{\beta,{\cal{A}}} \begin{array}{*{20}{c}}
{}\\
{}
\end{array}} \right\},
\label{inter-a-tilda}
\end{equation}
where $\tilde {\cal S}_{1,j}^{\cal A}=\bar {\cal S}_1^{[i]}$ in (\ref{set1}), $\tilde {\cal S}_{2,j}^{\cal A}=\tilde {\cal S}_2^{[i]} $ in (\ref{set5}), and $\tilde {\cal S}_{3,j}^{\cal A}=\tilde {\cal S}_3^{[i]} $ in (\ref{set6}).
In addition, we have:
\begin{equation}
\tilde {\cal S}_{{j'i'j}}^{\alpha,{\cal{A}}}= {\begin{cases}
{\left\{ {1,...,n} \right\},j' \ne j}\\
{\left\{ {1,...,n + 1} \right\},j' = j}
\end{cases}} ,
\end{equation}
\begin{equation}
\tilde {\cal S}_{{j'i'j}}^{\gamma,{\cal{A}}}=\{1,...,sn\},
\end{equation}
\begin{equation}
\tilde {\cal S}_{{j'i'j}}^{\beta,{\cal{A}}}=\{1,...,\upsilon n\}.
\end{equation}

By the same argument given for the clean receivers, subspaces $\bar {\cal {A}}_j$, $\tilde {\cal A}_j$ and $\bar {\cal C}_j$   are full rank  and linearly independent almost surely, i.e., all base vectors of these subspaces are linearly independent.
Now, we analyze the dimensions of the message and interference  subspaces. By calculating the number of base vectors of the message subspace $\bar {\cal C}_j$ in (\ref{c_bar_dirty}), we have:
\begin{equation}
d({\bar {\cal C}_j})={n^{{K^2} - K}}{(tn)^{QK}},
\label{d_2_1}
\end{equation}
\begin{equation*}
D_N ({\bar {\cal C}_j})=\zeta ,
\end{equation*}
and for the interference subspaces in (\ref{a_bar_dirty}) and (\ref{inter-a-tilda}), we have:
\begin{equation}
d({\bar {\cal A}_j})={n^{{K^2} - K - (K - 1)}}{(n + 1)^{K - 1}}{\left( {\max \{ sn,tn\} } \right)^{QK}} ,
\end{equation}
\begin{equation*}
D_N({\bar {\cal A}_j})=\max\{\Gamma,\zeta\},
\end{equation*}
\begin{equation}
d({\tilde {\cal A}_j})= {n^{{K^2} - K - (K - 1)}}{(n + 1)^{K - 1}}{(sn)^{QK - W}}{(\upsilon n)^W},
\label{d_2_3}
\end{equation}
\begin{equation*}
D_N({\tilde {\cal A}_j})=\chi.
\end{equation*}
Therefore, we can  see that the  interference alignment equations (\ref{IA3})-(\ref{IA6}) are satisfied, because the normalized  asymptotic dimension of  the interference subspace induced by ${\tilde {\bold V}}^{[i]} {\tilde {\bold x}}^{[i]},i\in \{1,..,W\}$ is $\chi$, the normalized asymptotic dimension of  the interference subspace induced by ${\bar {\bold V}}^{[i]} {\bar {\bold x}}^{[i]},i\in \{1,..,W\}$ is $\Gamma$, and  the normalized asymptotic dimension of  the interference subspace  induced by ${\bar {\bold V}}^{[i]} {\bar {\bold x}}^{[i]},i\in \{W+1,..,K\},i\ne j$ is $\zeta$.
\section{Proof of Lemma \ref{lemma_3}}
\label{appendix3}

Using (\ref{beam2}), we characterize the message  subspaces $\tilde {\cal{C}}_{i,r_q},i\in {\cal B}_q$ as follows:

\begin{fleqn}
\begin{equation*}
{\tilde {\cal C}_{i,r_q}} = \textrm{span}\left( {{{{\bold H}}_{\rm T-IR}^{[qi]}}{{\tilde {\bold V}}^{[i]}}} \right)=
\end{equation*}
\end{fleqn}
\begin{equation*}
\textrm{span} \left\{ {\left[ {\prod\limits_{(i',j') \in \tilde {\cal S}_{1,i,r_q}^{\cal C}} {{{\left( {{{\tilde{\bold H}}^{[j'i']}}} \right)}^{{\alpha _{j'i'}}}}} } \right]\left[ {\prod\limits_{(i'',q') \in \tilde {\cal S}_{2,i,r_q}^{\cal C}} {{{\left( {{{\bold H}_{\rm T-IR}^{[{{q'}}i'']}}} \right)}^{{\gamma _{{{q'}}i''}}}}} } \right]\left[ {\prod\limits_{(i''',q'') \in \tilde {\cal S}_{3,i,r_q}^{\cal C}} {{{\left( {{{\bold T}^{[{{q''}}i''']}}} \right)}^{{\beta _{{{q''}}i'''}}}}} } \right]} \right.{\bf w}:
\end{equation*}
\begin{equation}
\left. {{\alpha _{j'i'}} \in \tilde {\cal S}_{{j'i'ir_q}}^{\alpha ,{\cal C}},{\gamma _{{{q'}}i''}} \in \tilde {\cal S}_{{r_{q'}}i''ir_q}^{\gamma ,{\cal C}},{\beta _{{{q''}}i'''}} \in \tilde {\cal S}_{{r_{q''}}i'''ir_q}^{\beta ,{\cal C}}\begin{array}{*{20}{c}}
{}\\
{}
\end{array}} \right\},
\label{c_tilde_IR}
\end{equation}
where $\tilde {\cal S}_{1,i,r_q}^{\cal C}=\bar {\cal S}_1^{[i]}$ in (\ref{set1}), $\tilde {\cal S}_{2,i,r_q}^{\cal C}=\bar {\cal S}_2^{[i]} $ in (\ref{set2}), and $\tilde {\cal S}_{3,i,r_q}^{\cal C}=\tilde {\cal S}_3^{[i]}$ in (\ref{set6}). $\tilde {\cal S}_{{j'i'ir_q}}^{\alpha ,{\cal C}}$, $\tilde {\cal S}_{{r_{q'}}i''ir_q}^{\gamma ,{\cal C}}$, and $\tilde {\cal S}_{{r_{q''}}i'''ir_q}^{\beta ,{\cal C}}$ are defined as follows:
\begin{equation}
\tilde {\cal S}_{{j'i'ir_q}}^{\alpha ,{\cal C}} = \{1,...,n\},
\end{equation}
\begin{equation}
\tilde {\cal S}_{{r_{q'}}i''ir_q}^{\gamma ,{\cal C}} = {\begin{cases}
{\{ 1,...,sn\} ,q' \ne q,i'' \in \{ 1,...,K\} ,i'' \notin {{\cal B}_{q'}}}\\
{\{ 0\} ,q' \ne q,i'' \in {{\cal B}_{q'}}}\\
{\{ 1,...,sn\} ,q' = q,i'' \in \{ 1,...,K\} ,i'' \notin {{\cal B}_{q'}}}\\
{\{ 1\} ,q' = q,i'' = i}\\
{\{ 0\} ,q' = q,i'' \ne i,i\in{\cal B}_{q'}}
\end{cases}} \,
\end{equation}
\begin{equation}
\tilde {\cal S}_{{r_{q''}}i'''ir_q}^{\beta ,{\cal C}} = \left\{ {1,...,\upsilon n} \right\}.
\end{equation}

To satisfy the interference alignment equation (\ref{IA7}), the subspace $\bar{\cal A}_{r_q}$ must be chosen such that:
\begin{equation*}
\bigcup\limits_{i \in \{ 1,...,K\} } {\left\{ {{\text{span}}\left( {{\mathbf{H}}_{{\rm{T-IR}}}^{[qi]}{{\bar {\mathbf{V}} }^{[i]}}} \right)} \right\}}  \subseteq {\bar A _{{r_q}}}.
\end{equation*}
Therefore, we characterize ${{\bar {\cal{A}}}_j} $ as follows:
\begin{equation}
{{\bar {\cal{A}}}_{r_q}} =\textrm{span} \left\{ {\left[ {\prod\limits_{(i',j') \in \bar {\cal S}_{1,r_q}^{\cal{A}}} {{{\left( {{{\bold H}^{[j'i']}}} \right)}^{{\alpha _{j'i'}}}}} } \right]\left[ {\prod\limits_{(i'',q') \in \bar {\cal S}_{2,r_q}^{\cal{A}}} {{{\left( {{{\bold H}_{\rm T-IR}^{[{{q'}}i'']}}} \right)}^{{\gamma _{{{q'}}i''}}}}} } \right]{\bf w}:{\alpha _{j'i'}} \in \bar {\cal S}_{{j'i'r_q}}^{\alpha,{\cal{A}}} ,{\gamma _{{{q'}}i''}} \in \bar {\cal S}_{{r_{q'}}i''r_q}^{\gamma,{\cal{A}}} } \right\}
\label{a_bar_IR}
\end{equation}
where $\bar {\cal S}_{1,r_q}^{\cal{A}}=\bar {\cal S}_1^{[i]}$ in (\ref{set1}) and $\bar {\cal S}_{2,r_q}^{\cal{A}}=\bar {\cal S}_2^{[i]}$ in (\ref{set2}). $\bar {\cal S}_{{j'i'r_q}}^{\alpha,{\cal{A}}} $ and $\bar {\cal S}_{{r_{q'}}i''r_q}^{\gamma,{\cal{A}}}  $ are defined as follows: 
\begin{equation}
\bar {\cal S}_{{j'i'r_q}}^{\alpha,{\cal{A}}}  = \{1,...,n\},
\end{equation}
\begin{equation}
\bar {\cal S}_{{r_{q'}}i''r_q}^{\gamma,{\cal{A}}}  ={\begin{cases}
{\left\{ {1,...,\max \{ sn + 1,tn\} } \right\},q' = q,i'' \in \{ 1,...,W\} }\\
{\left\{ {1,...,\max \{ sn,tn + 1\} } \right\},q' = q,i'' \in \{ W + 1,...,K\} }\\
{\left\{ {1,...,\max \{ sn,tn\} } \right\},q' \ne q}
\end{cases}} .
\end{equation}

To satisfy the interference alignment equation (\ref{IA9}), the subspace $\tilde{\cal A}_{r_q}$ must be chosen such that:
\begin{equation*}
\bigcup\limits_{i \in \{ 1,...,W\},i\notin{\cal B}_q } {\left\{ {{\text{span}}\left( {{\mathbf{H}}_{{\rm{T-IR}}}^{[qi]}{{\tilde {\mathbf{V}} }^{[i]}}} \right)} \right\}}  \subseteq {\tilde A _{{r_q}}}.
\end{equation*}
Therefore, we characterize ${{\tilde {\cal{A}}}_j} $ as follows:
\begin{equation*}
\tilde {\cal A}_{r_q} =\textrm{span} \left\{ {\left[ {\prod\limits_{(i',j')\in\tilde {\cal S}_{1,r_q}^{\cal A}} {{{\left( {{{{\tilde{\bf H}}}^{[j'i']}}} \right)}^{{\alpha _{j'i'}}}}} } \right]\left[ {\prod\limits_{(i'',q')\in\tilde {\cal S}_{2,r_q}^{\cal A}} {{{\left( {{{\bf{H}}_{\rm T-IR}^{[{{q'}}i'']}}} \right)}^{{\gamma _{{{q'}}i''}}}}} } \right]\left[ {\prod\limits_{(i''',q'')\in\tilde {\cal S}_{3,r_q}^{\cal A}} {{{\left( {{\bold{T}^{[{{q''}}i''']}}} \right)}^{{\beta _{{{q''}}i'''}}}}} } \right]{\bf w}:} \right.
\end{equation*}
\begin{equation}
\left. {{\alpha _{j'i'}} \in \tilde {\cal S}_{{j'i'r_q}}^{\alpha,{\cal{A}}} ,{\gamma _{{{q'}}i''}} \in \tilde {\cal S}_{{j'i'r_q}}^{\gamma,{\cal{A}}} ,{\beta _{{{q''}}i'''}} \in \tilde {\cal S}_{{j'i'r_q}}^{\beta,{\cal{A}}} \begin{array}{*{20}{c}}
{}\\
{}
\end{array}} \right\},
\end{equation}
where $\tilde {\cal S}_{1,r_q}^{\cal A} =\bar {\cal S}_1^{[i]}$ in (\ref{set1}), $\tilde {\cal S}_{2,r_q}^{\cal A}=\tilde {\cal S}_2^{[i]}$ in (\ref{set5}), and $\tilde {\cal S}_{3,r_q}^{\cal A}=\tilde {\cal S}_3^{[i]}$ in (\ref{set6}). In addition, we have:
\begin{equation}
\tilde {\cal S}_{{j'i'r_q}}^{\alpha,{\cal{A}}}=\{1,...,n\},
\end{equation}
\begin{equation}
\tilde {\cal S}_{{j'i'r_q}}^{\gamma,{\cal{A}}}= {\begin{cases}
{\left\{ {1,...,sn} \right\},q' \ne q,i'' \in \left\{ {1,...,K} \right\},i'' \notin {{\cal B}_{q'}}}\\
{\left\{ {1,...,sn + 1} \right\},q' = q,i'' \in \left\{ {1,...,W} \right\},i'' \notin {{\cal B}_{q'}}}\\
{\left\{ {1,...,sn} \right\},q' = q,i'' \in \left\{ {W + 1,...,K} \right\}}
\end{cases}} ,
\end{equation}
\begin{equation}
\tilde {\cal S}_{{j'i'r_q}}^{\beta,{\cal{A}}}=\{1,...,\upsilon n\}.
\end{equation}

By the same argument given for clean receivers, subspaces $\bar {\cal {A}}_{r_q}$, $\tilde {\cal A}_{r_q}$ and $\tilde {\cal C}_{i,r_q},i\in {\cal B}_q$   are full rank  and linearly independent almost surely, i.e., all base vectors of these subspaces are linearly independent.
Now, by calculating the number of base vectors, we analyze the dimensions of the subspaces  $\tilde {\cal{C}}_{i,r_q},i\in { B}_q$, $\bar{\cal{A}}_{r_q}$ and $\tilde{\cal{A}}_{r_q}$:
\begin{equation}
d(\tilde {\cal{C}}_{i,r_q})={n^{{K^2} - K}}{(sn)^{QK - W}}{(\upsilon n)^W},\forall i \in {\cal B}_q,
\label{d_3_1}
\end{equation}
\begin{equation*}
D_N(\tilde {\cal{C}}_{i,r_q})=\chi.
\end{equation*}
So, the normalized dimension of the total subspaces whose message symbols could be demultiplexed ($\tilde{\bold x}^{[i]},i\in {\cal B}_q$) at the MIMO C-IR $q$-th receive antenna is:
\begin{equation*}
\sum\limits_{i \in {{\cal B}_q}} {D_ N(\tilde {\cal{C}}_{i,r_q})}= \left| {{{\cal B}_q}} \right|\chi  .
\end{equation*}

For $\bar {\cal{A}}_{r_q}$, we have:
\begin{equation}
d(\bar {\cal{A}}_{r_q})={n^{{K^2} - K}}{\left( {\max \{ sn,tn\} } \right)^{K(Q - 1)}}{\left( {\max \{ sn + 1,tn\} } \right)^W}{\left( {\max \{ sn,tn + 1\} } \right)^{K - W}},
\end{equation}
\begin{equation*}
D_N(\bar {\cal{A}}_{r_q})=\max\{\Gamma,\zeta\},
\end{equation*}
and for $\tilde {\cal{A}}_{r_q}$, we have:
\begin{equation}
d(\tilde {\cal{A}}_{r_q})={n^{{K^2} - K}}{(sn)^{QK - W - (W - \left| {{{\cal B}_q}} \right|)}}{(sn + 1)^{W - \left| {{{\cal B}_q}} \right|}}{(\upsilon n)^W},
\label{d_3_3}
\end{equation}
\begin{equation*}
D_N(\tilde {\cal{A}}_{r_q})=\chi.
\end{equation*}
Thus, we can  see that interference alignment equations (\ref{IA7})-(\ref{IA10}) are satisfied. 

\section{Proof of Theorem \ref{theorem2}}
\label{appendix5}
The second term of (\ref{DoF2}) is exactly the same as the second term of (\ref{DoF1}) in Theorem \ref{theorem1}. The proof of the first term is similar to the proof of the first term of (\ref{DoF1}) in  Theorem \ref{theorem1} with a difference in the MIMO C-IR demultiplexing method. In the proof of  Theorem \ref{theorem1}, each MIMO C-IR receive antenna $q$, demultiplexes the message streams $\tilde {\bold x}_i,i\in {\cal B}_q$ separately without coordination with other receive antennas. However, in the proof of this theorem, we use coordination between MIMO C-IR receive antennas. Without loss of generality, assume that $\left| {{{\cal B}_1}} \right| = Z + 1$ and $\left| {{{\cal B}_q}} \right| = Z,q \ne 1$. To demultiplex the message streams  $\tilde {\bold x}_i,i\in\{1,...,W\}$ at the MIMO C-IR, first we  demultiplex the message streams  $\tilde {\bold x}_i,i\in {\cal B}_q,q\ne1$ at the $q$-th MIMO C-IR receive antenna separately. Then, to demultiplex the message streams  $\tilde {\bold x}_i,i\in {\cal B}_1$, we first remove the interference induced by the message streams  $\tilde {\bold x}_i,i\in \{1,...,W\},i\notin {\cal B}_1$. This results in a decrement in the total normalized asymptotic dimension at the first receive antenna of the MIMO C-IR   (the amount of decrement is $\chi$), so (\ref{D_{N,t,r_q}}) changes into the following form for $q=1$:
\begin{equation}
D_{N,t,r_1}=\left\lfloor {\frac{W}{Q}} \right\rfloor  \chi +\chi+\max\{\Gamma,\zeta\},
\end{equation}
and the constraint (\ref{assumption2}) changes into the following form:
\begin{equation}
\zeta  \ge \left\lfloor {\frac{W}{Q}} \right\rfloor \chi.
\end{equation}
Then, we see that the DoF (\ref{DoF2}) is achievable.

\section{Proof of Theorem \ref{theorem3}}
\label{appendix6}
The proof of this theorem is similar to the first term in the proof of Theorem \ref{theorem1}.
Here, we  use the variable $U$ introduced in the statement of the theorem to denote the number of clean receivers. Note that to avoid several notations, we use the same notations (such as the name of sets and vector subspaces) used in the proof of Theorem \ref{theorem1}. So from now on, these notations belong to this theorem. Our proof has six steps as follows:

\textbf{Step 1: Dividing receivers, transmitters, and NC-IR antennas}

The same as Step 1 of the proof of first term in Theorem \ref{theorem1}, we divide the transmitters into two partitions. For the transmitters $ i \in \{1,...,U\}$, we provide two sets of symbol streams ${{\bar {\bold{x}}}^{[i]}}$ and ${{\tilde {\bold{x}}}^{[i]}}$. The matrices ${{\bar{\bold{V}}}^{[i]}}$ and  ${{\tilde{\bold{V}}}^{[i]}}$ are the beamforming matrices, whose columns are the beamforming vectors for each element of ${{\bar {\bold{x}}}^{[i]}}$ and ${{\tilde{\bold{x}}}^{[i]}}$, respectively.
For the transmitters $i\in \{U+1,...,K\}$,  we only provide one set of  symbol stream ${{\bar {\bold{x}}}^{[i]}}$ and matrix ${{\bar{\bold{V}}}^{[i]}}$ is the beamforming matrix for the symbols ${{\bar {\bold{x}}}^{[i]}}$. Hence, the vectors ${\bold{X}^{[i]}} $ will have the form of  (\ref{partition1}) and (\ref{partition2}) by setting $W=U$. The reason of this kind of partitioning is the same as Theorem \ref{theorem1}. The main difference here is in the interference alignment scheme used for demultiplexing  the message streams ${{\tilde {\bold{x}}}^{[i]}},i\in \{1,...,U\}$ in 
the NC-IR receive antennas.

Next, we divide the transmitters $i\in \{1,...,U\}$ into  $p$ distinct sets ${\cal E}_l,l\in \{1,...,p\}$ such that for $l\in \{1,...,e'\}$, we have $\left| {{{\cal E}_l}} \right| = e + 1$  and for $l\in \{e'+1,...,p\}$, we have $\left| {{{\cal E}_l}} \right| = e$. Similarly, we divide the NC-IR antennas into $p$ distinct sets ${\cal F}_l,l\in \{1,...,p\}$ such that $\left| {{{\cal F}_l}} \right| = U,\forall l \in \{1,...,p\}$. Now, we design the beamforming matrices ${{\bar{\bold{V}}}^{[i]}}$ and  ${{\tilde{\bold{V}}}^{[i]}}$ such that the message streams ${{\tilde{\bold{x}}}^{[i]}}, i \in {\cal E}_l$ could be demultiplexed in each of the NC-IR antenna $u \in {\cal F}_l$ for $\forall l \in \{1,...,p\}$.

\vspace{10pt}
\textbf{Step 2: Interference cancellation at the clean receivers and equivalent channel at the dirty receivers}

For the interference cancellation, we design the outputs of antennas in the set ${\cal F}_l$ such that the interference induced by the message streams  ${{\tilde {\bold{x}}}^{[i]}},i\in {\cal E}_l$ is removed at the clean receivers $j\in \{1,...,U\}$. So, the NC-IR antennas transmitted signal must be designed such that they satisfy the following:
\begin{equation}
 - \sum\limits_{i \in {{\cal E}_l},i \ne j} {{{\bold H}^{[ji]}}{{\tilde {\bold V}}^{[i]}}{{\tilde {\bold x}}^{[i]}}}  = \sum\limits_{u \in {{\cal F}_l}} {{{\bold H}_{\rm IR-R}^{[j{u}]}}{{\bold X}_{\rm IR}^{[{u}]}}} ,\forall j \in \left\{ {1,...,U} \right\},\forall l \in \left\{ {1,...,p} \right\}.
\label{interference cancellation4}
\end{equation}
 The solution of (\ref{interference cancellation4}) can be derived as follows:
\begin{equation}
{{\bold X}_{\rm IR}^{[{u}]}} = \sum\limits_{j \in \{ 1,...,U\} } {\sum\limits_{i \in {{\cal E}_l},i \ne j} {{\bold H}_{\rm inv}^{[j{u}]}{{\bold H}^{[ji]}}{{\tilde {\bold V}}^{[ji]}}{{\tilde {\bold x}}^{[i]}},\forall u \in {{\cal F}_l}} },
\label{NC-IR antennas}
\end{equation}
where ${\bold H}_{\rm inv}^{[j{u}]}$ is a $T\times T$ diagonal matrix and its $t$-th diagonal element is a fractional polynomial in terms of ${ H}_{\rm IR-R}^{[j'{u'}]}(\omega_t),u'\in {\cal F}_l,j'\in\{1,...,U\}$. This solution exists almost surely, because the matrix of coefficients of the linear equations is in terms of independent random variables and its determinant is a non-zero polynomial in terms of these random variables drawn from a continuous cumulative probability distribution, and by \cite[Lemma 1]{me}, it is non-zero with probability $1$.
 Note that each NC-IR receive antenna demultiplexes  symbol streams $\tilde{\bf x}^{[i]}$ with an additive noise. This event does not disturb the equations above, because if each symbol is replaced by a symbol with an additive noise, the interference cancellation holds, but we will have an additional noise, which is negligible in high SNR regime.
We can see that the received signals at the receivers have the same form as (\ref{receivers}) and ${\bold H}_{\rm inv}^{[j{u}]}$ and equivalent channel matrix ${{{\tilde{\bf H}}}^{[ji]}}$ have the same properties introduced in Step 2 of the proof of the first term in Theorem \ref{theorem1}.

\vspace{30pt}
\textbf{Step 3: Interference alignment equations}

Interference alignment equations and message and interference subspaces for the clean and dirty receivers are the same as Step 3 in the proof of first term in Theorem \ref{theorem1} ((\ref{IA1})-(\ref{IA6})), if we replace $W$ with $U$. 
Consider a $q\in \{1,...,pU\}$: we define the function $L(q)=l$ if $q\in {\cal F}_l$ ($l$ is unique because the sets ${\cal F}_l$ are disjoint). We will design the interference alignment scheme such that the symbol streams $\tilde{\bf x}^{[i]},i\in {\cal E}_{L(q)}$ can be demultiplexed at the $q$-th receive antenna of the NC-IR.
Thus, interference alignment equations for the NC-IR will change as follows:

To this end, all the interference induced by the symbol streams $\bar{\bf x}^{[i]}$ must align into a limited subspace. Therefore, at the $q$-th receive  antenna of the NC-IR and for each $i \in \{1,...,K\}$, we must have:
\begin{equation}
\textrm{span}\left( {{\bold{H}_{\rm T-IR}^{[qi]}}{{\bar {\bold V}}^{[i]}}} \right) \subseteq {\bar{\cal A}_{r_q}},
\label{IA11}
\end{equation}
where ${\bar{\cal A}_{r_q}}$ is considered as a subspace, for which we have:
\begin{equation}
\mathop {\max }\limits_{i\in \{1,...,K\}} {D_N}\left(\textrm{span}\left({\bold{H}_{\rm T-IR}^{[qi]}}{{\bar {\bold V}}^{[i]}}\right)\right) = {D_N}({\bar{\cal A}_{r_q}}).
\label{IA12}
\end{equation}
Then, for each $i \in \{1,...,U\},i \notin {\cal E}_{L(q)}$, we have:
\begin{equation}
\textrm{span}\left( {{\bold{H}_{\rm T-IR}^{[qi]}}{{\tilde {\bold V}}^{[i]}}} \right) \subseteq {\tilde{\cal A}_{r_q}},
\label{IA13}
\end{equation}
where ${\tilde{\cal A}_{r_q}}$ is considered as a subspace, for which we have:
\begin{equation}
\mathop {\max }\limits_{i \in \{1,...,U\},i  \notin  {\cal E}_{L(q)}} {D_N}\left(\textrm{span}\left({\bold{H}_{\rm T-IR}^{[qi]}}{{\tilde {\bold V}}^{[i]}}\right)\right) = {D_N}({\tilde{\cal A}_{r_q}}).
\label{IA14}
\end{equation}
Also, we define ${\tilde {\cal C}_{i,r_q}},i\in {\cal E}_{L(q)}$ as the  message subspaces, which can be demultiplexed at the NC-IR $q$-th antenna as follows:
\begin{equation}
{\tilde {\cal C}_{i,r_q}} = \textrm{span}\left( {{{{\bold H}}_{\rm T-IR}^{[qi]}}{{\tilde {\bold V}}^{[i]}}} \right),i \in {\cal E}_{L(q)},
\end{equation}
and we want ${\tilde {\cal C}_{i,r_q}}, \forall i\in {\cal E}_{L(q)}$, $\bar {\cal A}_{r_q}$ and $\tilde {\cal A}_{r_q}$ to be full rank and linearly independent,  so we can make sure that  the message streams $\tilde {\bold x}_i,i\in {\cal E}_{L(q)}$ could be demultiplexed at the $q$-th NC-IR antenna. In Steps 4 and 5, we prove the existence of such beamforming vectors, message, and interference subspaces, which satisfy the previous interference alignment equations for the clean and dirty receivers and the MIMO C-IR. In Step 6, we analyze the achieved DoF by these beamforming vectors design.

\vspace{10pt}

\textbf{Step 4: Beamforming matrix design}

The beamforming matrices ${{\bar{\bold{V}}}^{[i]}}, \forall i \in \{1,...,K\}$, is the same as (\ref{beam1}) and (\ref{beam3}), if we replace $W$ with $U$.
For ${{\tilde{\bold{V}}}^{[i]}}$, we have:
\begin{equation}
{{{\tilde{\bf{ V}}}}^{[i]}} = \left\{ {\left[ {\prod\limits_{(i',j')\in\tilde {\cal S}_1^{[i]}} {{{\left( {{{{\tilde{\bf H}}}^{[j'i']}}} \right)}^{{\alpha _{j'i'}}}}} } \right]\left[ {\prod\limits_{(i'',q')\in\tilde {\cal S}_2^{[i]}} {{{\left( {{{\bf{H}}_{\rm T-IR}^{[{{q'}}i'']}}} \right)}^{{\gamma _{{{q'}}i''}}}}} } \right]\left[ {\prod\limits_{(i''',q'')\in\tilde {\cal S}_3^{[i]}} {{{\left( {{\bold{T}^{[{{q''}}i''']}}} \right)}^{{\beta _{{{q''}}i'''}}}}} } \right]{\bf w}:} \right.
\end{equation}
\begin{equation}
\left. {{\alpha _{j'i'}} \in \{ 1,...,n\} ,{\gamma _{{{q'}}i''}} \in \{ 1,...,sn\} ,{\beta _{{{q''}}i'''}} \in \{ 1,...,\upsilon n\} \begin{array}{*{20}{c}}
{}\\
{}
\end{array}} \right\},
\end{equation}
where $\tilde {\cal S}_1^{[i]}=\bar {\cal S}_1^{[i]}$ in (\ref{set1}), and we have:
\begin{equation}
\tilde {\cal S}_2^{[i]} = \left\{ {(i'',q')\left| {i'' \in \{ 1,...,K\} ,i'' \notin {\cal E}_{L(q')},q' \in \{ 1,...,Q\} } \right.} \right\},
\label{set8}
\end{equation}
\begin{equation}
\tilde {\cal S}_3^{[i]} = \left\{ {(i''',q'')\left| {i''' \in {\cal E}_{L(q'')},q'' \in \{ 1,...,Q\} } \right.} \right\},
\label{set9}
\end{equation}
and ${{\bold{T}^{[{{q''}}i''']}}}$s  are $T\times T$ diagonal random matrices for each $(i''',q'')$, where each of  diagonal elements for each matrix  is drawn independently from  a continuous cumulative probability distribution.

Note that similar to the proof of Theorem \ref{theorem1},  each value of parameters $s,\upsilon$ and $t$ can be approximated by rational numbers with arbitrarily small error, and by choosing a sufficiently large $n$, parameters $sn,\upsilon n$ and $tn$ will be integers.

\vspace{10pt}

\textbf{Step 5: Validity of interference alignment conditions and decodability of message symbols}

\vspace{10pt}

\textit{1) Validity of interference alignment conditions at clean receivers $j\in \{1,...,U\}$:}

The message subspace $\bar{\cal{C}}_j$ and the interference subspace $\bar{\cal{A}}_j$ will be the exactly same as (\ref{c_bar_clean}) and (\ref{a_bar_clean}).
The message  subspaces $\tilde{\cal{C}}_j$ will change as follows:

\begin{fleqn}
\begin{equation*}
{\tilde {\cal C}_j} = \textrm{span}\left( {{{\tilde{\bold H}}^{[jj]}}{{\tilde {\bold V}}^{[j]}}} \right)=
\end{equation*}
\end{fleqn}
\begin{equation*}
 \textrm{span} \left\{ {\left[ {\prod\limits_{(i',j') \in \tilde {\cal S}_{1,j}^{\cal C}} {{{\left( {{{\tilde{\bold H}}^{[j'i']}}} \right)}^{{\alpha _{j'i'}}}}} } \right]\left[ {\prod\limits_{(i'',q') \in \tilde {\cal S}_{2,j}^{\cal C}} {{{\left( {{{\bold H}_{\rm T-IR}^{[{{q'}}i'']}}} \right)}^{{\gamma _{{{q'}}i''}}}}} } \right]\left[ {\prod\limits_{(i''',q'') \in \tilde {\cal S}_{3,j}^{\cal C}} {{{\left( {{{\bold T}^{[{{q''}}i''']}}} \right)}^{{\beta _{{{q''}}i'''}}}}} } \right]} \right.{\bf w}:
\end{equation*}
\begin{equation}
\left. {{\alpha _{j'i'}} \in \tilde {\cal S}_{{j'i'j}}^{\alpha ,{\cal C}},{\gamma _{{{q'}}i''}} \in \tilde {\cal S}_{{r_{q'}}i''j}^{\gamma ,{\cal C}},{\beta _{{{q''}}i'''}} \in \tilde {\cal S}_{{r_{q''}}i'''j}^{\beta ,{\cal C}}\begin{array}{*{20}{c}}
{}\\
{}
\end{array}} \right\},
\label{c_tilde_clean_NC-IR}
\end{equation}
where $\tilde {\cal S}_{1,j}^{\cal C}=\bar {\cal S}_{1,j}^{\cal C}$ in (\ref{set4}), $\tilde {\cal S}_{2,j}^{\cal C}=\tilde {\cal S}_2^{[i]}$ in (\ref{set8}), and $\tilde {\cal S}_{3,j}^{\cal C}=\tilde {\cal S}_3^{[i]}$ in (\ref{set9}). $\tilde {\cal S}_{{j'i'j}}^{\alpha ,{\cal C}}$, $\tilde {\cal S}_{{r_{q'}}i''j}^{\gamma ,{\cal C}}$, and $\tilde {\cal S}_{{r_{q''}}i'''j}^{\beta ,{\cal C}}$ are defined as follows:
\begin{equation}
\tilde {\cal S}_{j'i'j}^{\alpha ,{\cal C}} =  {\begin{cases}
{\{ 1,...,n\} ,i' \ne j'}\\
{\{ 0\} ,j' \ne j,i' = j'}\\
{\{ 1\} ,j' = j,i' = j}
\end{cases}} ,
\end{equation}
\begin{equation}
\tilde {\cal S}_{{r_{q'}}i''j}^{\gamma ,{\cal C}} = \left\{ {1,...,sn} \right\},
\end{equation}
\begin{equation}
\tilde {\cal S}_{{r_{q''}}i'''j}^{\beta ,{\cal C}} = \left\{ {1,...,\upsilon n} \right\}.
\end{equation}

By the nature of $\bar {\cal {A}}_j$ in (\ref{a_bar_clean}), $\bar {\cal C}_j$ in (\ref{c_bar_clean}) and $\tilde {\cal C}_j$ in (\ref{c_tilde_clean_NC-IR}), we can see from the statement of the \cite[Lemma 2]{me} that if we choose the variables $x_k$ as $H^{[ji]}(\omega_t),H_{\rm T-IR}^{[ri']}(\omega_t),i,i',j\in \{1,...,K\},u\in\{1,...,Q\}$, $y_k$ as $H_{\rm IR-R}^{[ju]}(\omega_t),j\in \{U+1,...,K\},u\in \{1,...,Q\}$ and $z_k$ as $H_{\rm IR-R}^{[ju]}(\omega_t),j\in \{1,...,U\},u\in \{1,...,Q\}$, then by \cite[Lemmas1-3]{me},  subspaces $\bar {\cal {A}}_j$, $\bar {\cal C}_j$, and $\tilde {\cal C}_j$ are full rank  and linearly independent  (all base vectors of these subspaces are linearly independent) almost surely. The reason is that if we take the constructing base vectors of $\bar {\cal {A}}_j$, $\bar {\cal C}_j$ and $\tilde {\cal C}_j$ and construct a square matrix by choosing some rows of it, we can see by \cite[Lemmas 2-3]{me} that the determinant of this square matrix is a non-zero polynomial, which is non-zero with probability $1$ by \cite[Lemma 1]{me}. Thus, all message streams are decodable at the clean receivers (by zero forcing).
For more clarity, \cite[Lemmas 1-3]{me} have been reviewed in Appendix \ref{appendix1}.

Similar to the proof of Theorem \ref{theorem1}, first we assume that the parameter $T$ is sufficiently large and at the end of this step, we determine the minimum required $T$.
The dimensions of the subspaces  $\bar {\cal C}_j$ and $\bar {\cal A}_j$ are the same as (\ref{dim_mess_clean1}) and (\ref{dim_inter_clean}), respectively. Hence, we calculate the dimension of $\tilde {\cal C}_j$ by calculating the number of its base vectors in (\ref{c_tilde_clean_NC-IR}) as follows:
\begin{equation}
d(\tilde {\cal C}_j) ={n^{{K^2} - K}}{(sn)^\varphi }{(\upsilon n)^\theta },
\label{dim_mess_clean2}
\end{equation}
where
\begin{equation*}
\varphi  = \sum\limits_{q' = 1}^Q {(K - \left| {{{\cal E}_{L(q')}}} \right|)}  = KQ - \sum\limits_{q' = 1}^Q {\left| {{{\cal E}_{L(q')}}} \right|}  = KQ - U^2,
\end{equation*}
\begin{equation*}
\theta  = \sum\limits_{q' = 1}^Q {\left| {{{\cal E}_{L(q')}}} \right|}  = U^2.
\end{equation*}

We can see  from (\ref{DN}) that $l=K^2-K+QK$.
We define the following parameters:
\begin{equation*}
\Gamma  = {s^{QK}},
\end{equation*}
\begin{equation*}
\chi  = {s^{QK-U^2}}{\upsilon ^{U^2}},
\end{equation*}
\begin{equation*}
\zeta  = {t^{QK}}.
\end{equation*}
Therefore, the normalized asymptotic dimensions of the message and interference subspaces are:
\begin{equation}
D_N(\bar {\cal C}_j)=\Gamma,
\end{equation}
\begin{equation}
D_N(\tilde {\cal C}_j)=\chi,
\end{equation}
\begin{equation}
D_N({\bar{\cal A}}_j)=\max\{\Gamma,\zeta\}.
\end{equation}
Thus, the  interference alignment equations (\ref{IA1}) and (\ref{IA2}) are satisfied.

\vspace{10pt}

\textit{2) Validity of interference alignment conditions at dirty receivers $j\in \{U+1,...,K\}$:}


For dirty receivers, the message subspace $\bar{\cal{C}}_j$ and the interference subspace $\bar{\cal{A}}_j$ are exactly the same as (\ref{c_bar_dirty}) and (\ref{a_bar_dirty}).
To satisfy the interference alignment equation (\ref{IA5}) (if $W$ is replaced with $U$), the subspace $\tilde{\cal A}_j$ must be chosen such that:

\begin{equation*}
\bigcup\limits_{i\in\{1,...,U\}} {\left\{ {{\text{span}}\left( {{\tilde{\mathbf{H}}^{[ji]}}{{\tilde {\mathbf{V}} }^{[i]}}} \right)} \right\}}  \subseteq {\tilde {\cal A} _j}.
\end{equation*}
Therefore, we characterize subspace ${{\tilde {\cal{A}}}_j} $ as follows:\begin{equation*}
\tilde {\cal A}_j =\textrm{span} \left\{ {\left[ {\prod\limits_{(i',j')\in\tilde {\cal S}_{1,j}^{\cal A}} {{{\left( {{{{\tilde{\bf H}}}^{[j'i']}}} \right)}^{{\alpha _{j'i'}}}}} } \right]\left[ {\prod\limits_{(i'',q')\in\tilde {\cal S}_{2,j}^{\cal A}} {{{\left( {{{\bf{H}}_{\rm T-IR}^{[{{q'}}i'']}}} \right)}^{{\gamma _{{{q'}}i''}}}}} } \right]\left[ {\prod\limits_{(i''',q'')\in\tilde {\cal S}_{3,j}^{\cal A}} {{{\left( {{\bold{T}^{[{{q''}}i''']}}} \right)}^{{\beta _{{{q''}}i'''}}}}} } \right]{\bf w}:} \right.
\end{equation*}
\begin{equation}
\left. {{\alpha _{j'i'}} \in \tilde {\cal S}_{{j'i'j}}^{\alpha,{\cal{A}}} ,{\gamma _{{{q'}}i''}} \in \tilde {\cal S}_{{j'i'j}}^{\gamma,{\cal{A}}} ,{\beta _{{{q''}}i'''}} \in \tilde {\cal S}_{{j'i'j}}^{\beta,{\cal{A}}} \begin{array}{*{20}{c}}
{}\\
{}
\end{array}} \right\},
\end{equation}
where $\tilde {\cal S}_{1,j}^{\cal A}=\bar {\cal S}_1^{[i]}$ in (\ref{set1}), $\tilde {\cal S}_{2,j}^{\cal A}=\tilde {\cal S}_2^{[i]} $ in (\ref{set8}), and $\tilde {\cal S}_{3,j}^{\cal A}=\tilde {\cal S}_3^{[i]} $ in (\ref{set9}). In addition, we have:
\begin{equation}
\tilde {\cal S}_{{j'i'j}}^{\alpha,{\cal{A}}}= {\begin{cases}
{\left\{ {1,...,n} \right\},j' \ne j}\\
{\left\{ {1,...,n + 1} \right\},j' = j}
\end{cases}} ,
\end{equation}
\begin{equation}
\tilde {\cal S}_{{j'i'j}}^{\gamma,{\cal{A}}}=\{1,...,sn\},
\end{equation}
\begin{equation}
\tilde {\cal S}_{{j'i'j}}^{\beta,{\cal{A}}}=\{1,...,\upsilon n\}.
\end{equation}

By the same argument given for $\bar {\cal {A}}_j$, $\bar {\cal C}_j$, and $\tilde {\cal C}_j$ at clean receivers, subspaces $\bar {\cal {A}}_j$, $\tilde {\cal A}_j$ and $\bar {\cal C}_j$   are full rank  and linearly independent almost surely.
Then, we have:
\begin{equation}
D_N ({\bar {\cal C}_j})=\zeta ,
\end{equation}
\begin{equation}
D_N({\bar {\cal A}_j})=\max\{\Gamma,\zeta\},
\end{equation}
\begin{equation}
d({\tilde {\cal A}_j})= {n^{{K^2} - K - (K - 1)}}{(n + 1)^{K - 1}}{(sn)^{QK - U^2}}{(\upsilon n)^{U^2}},
\end{equation}
\begin{equation}
D_N({\tilde {\cal A}_j})=\chi,
\end{equation}
hence, we can  see that the  interference alignment equations (\ref{IA3})-(\ref{IA6}) are satisfied.


\textit{3) Validity of interference alignment conditions at the  $q$-th antenna of the NC-IR $q\in \{1,...,Q\}$:}


The interference subspace $\bar{\cal A}_{r_q}$ is exactly the same as (\ref{a_bar_IR}), if we replace $W$ with $U$.
The  message  subspaces $\tilde {\cal{C}}_{i,r_q},i\in {\cal E}_{L(q)}$  and the interference subspace  $\tilde{\cal{A}}_{r_q}$ will change as follows:
\begin{fleqn}
\begin{equation*}
{\tilde {\cal C}_{i,r_q}} = \textrm{span}\left( {{{{\bold H}}_{\rm T-IR}^{[qi]}}{{\tilde {\bold V}}^{[i]}}} \right)=
\end{equation*}
\end{fleqn}
\begin{equation*}
 \textrm{span} \left\{ {\left[ {\prod\limits_{(i',j') \in \tilde {\cal S}_{1,i,r_q}^{\cal C}} {{{\left( {{{\tilde{\bold H}}^{[j'i']}}} \right)}^{{\alpha _{j'i'}}}}} } \right]\left[ {\prod\limits_{(i'',q') \in \tilde {\cal S}_{2,i,r_q}^{\cal C}} {{{\left( {{{\bold H}_{\rm T-IR}^{[{{q'}}i'']}}} \right)}^{{\gamma _{{{q'}}i''}}}}} } \right]\left[ {\prod\limits_{(i''',q'') \in \tilde {\cal S}_{3,i,r_q}^{\cal C}} {{{\left( {{{\bold T}^{[{{q''}}i''']}}} \right)}^{{\beta _{{{q''}}i'''}}}}} } \right]} \right.{\bf w}:
\end{equation*}
\begin{equation}
\left. {{\alpha _{j'i'}} \in \tilde {\cal S}_{{j'i'ir_q}}^{\alpha ,{\cal C}},{\gamma _{{{q'}}i''}} \in \tilde {\cal S}_{{r_{q'}}i''ir_q}^{\gamma ,{\cal C}},{\beta _{{{q''}}i'''}} \in \tilde {\cal S}_{{r_{q''}}i'''ir_q}^{\beta ,{\cal C}}\begin{array}{*{20}{c}}
{}\\
{}
\end{array}} \right\},
\label{c_tilde_NC-IR}
\end{equation}
where $\tilde {\cal S}_{1,i,r_q}^{\cal C}=\bar {\cal S}_1^{[i]}$ in (\ref{set1}), $\tilde {\cal S}_{2,i,r_q}^{\cal C}=\bar {\cal S}_2^{[i]}$ in (\ref{set2}), and $\tilde {\cal S}_{3,i,r_q}^{\cal C}=\tilde {\cal S}_3^{[i]}  $ in (\ref{set9}). $\tilde {\cal S}_{{j'i'ir_q}}^{\alpha ,{\cal C}}$, $\tilde {\cal S}_{{r_{q'}}i''ir_q}^{\gamma ,{\cal C}}$, and $\tilde {\cal S}_{{r_{q''}}i'''ir_q}^{\beta ,{\cal C}}$ are defined as follows:
\begin{equation}
\tilde {\cal S}_{{j'i'ir_q}}^{\alpha ,{\cal C}} = \{1,...,n\},
\end{equation}
\begin{equation}
\tilde {\cal S}_{{r_{q'}}i''ir_q}^{\gamma ,{\cal C}} = {\begin{cases}
{\{ 1,...,sn\} ,q' \ne q,i'' \in \{ 1,...,K\} ,i'' \notin {{\cal E}_{L(q')}}}\\
{\{ 0\} ,q' \ne q,i'' \in {{\cal E}_{L(q')}}}\\
{\{ 1,...,sn\} ,q' = q,i'' \in \{ 1,...,K\} ,i'' \notin {{\cal E}_{L(q')}}}\\
{\{ 1\} ,q' = q,i'' = i}\\
{\{ 0\} ,q' = q,i'' \ne i,{\color{black} i''\in{{\cal E}_{L(q')}}}}
\end{cases}} \,
\end{equation}
\begin{equation}
\tilde {\cal S}_{{r_{q''}}i'''ir_q}^{\beta ,{\cal C}} = \left\{ {1,...,\upsilon n} \right\}.
\end{equation}

To satisfy the interference alignment equation (\ref{IA13}), the subspace $\tilde{\cal A}_{r_q}$ must be chosen such that:
\begin{equation*}
\bigcup\limits_{i \in \{ 1,...,U\},i\notin{\cal E}_{L(q)} } {\left\{ {{\text{span}}\left( {{\mathbf{H}}_{{\rm{T-IR}}}^{[qi]}{{\tilde {\mathbf{V}} }^{[i]}}} \right)} \right\}}  \subseteq {\tilde A _{{r_q}}}.
\end{equation*}
Therefore, we characterize ${{\tilde {\cal{A}}}_j} $ as follows:
\begin{equation*}
\tilde {\cal A}_{r_q} =\textrm{span} \left\{ {\left[ {\prod\limits_{(i',j')\in\tilde {\cal S}_{1,r_q}^{\cal A}} {{{\left( {{{{\tilde{\bf H}}}^{[j'i']}}} \right)}^{{\alpha _{j'i'}}}}} } \right]\left[ {\prod\limits_{(i'',q')\in\tilde {\cal S}_{2,r_q}^{\cal A}} {{{\left( {{{\bf{H}}_{\rm T-IR}^{[{{q'}}i'']}}} \right)}^{{\gamma _{{{q'}}i''}}}}} } \right]\left[ {\prod\limits_{(i''',q'')\in\tilde {\cal S}_{3,r_q}^{\cal A}} {{{\left( {{\bold{T}^{[{{q''}}i''']}}} \right)}^{{\beta _{{{q''}}i'''}}}}} } \right]{\bf w}:} \right.
\end{equation*}
\begin{equation}
\left. {{\alpha _{j'i'}} \in \tilde {\cal S}_{{j'i'r_q}}^{\alpha,{\cal{A}}} ,{\gamma _{{{q'}}i''}} \in \tilde {\cal S}_{{j'i'r_q}}^{\gamma,{\cal{A}}} ,{\beta _{{{q''}}i'''}} \in \tilde {\cal S}_{{j'i'r_q}}^{\beta,{\cal{A}}} \begin{array}{*{20}{c}}
{}\\
{}
\end{array}} \right\},
\end{equation}
where $\tilde {\cal S}_{1,r_q}^{\cal A}=\bar {\cal S}_1^{[i]}$ in (\ref{set1}), $\tilde {\cal S}_{2,r_q}^{\cal A}=\tilde {\cal S}_2^{[i]}$ in (\ref{set8}), and $\tilde {\cal S}_{3,r_q}^{\cal A}=\tilde {\cal S}_3^{[i]} $ in (\ref{set9}). In addition, we have:
\begin{equation}
\tilde {\cal S}_{{j'i'r_q}}^{\alpha,{\cal{A}}}=\{1,...,n\},
\end{equation}
\begin{equation}
\tilde {\cal S}_{{j'i'r_q}}^{\gamma,{\cal{A}}}= {\begin{cases}
{\left\{ {1,...,sn} \right\},q' \ne q,i'' \in \left\{ {1,...,K} \right\},i'' \notin {{\cal E}_{L(q')}}}\\
{\left\{ {1,...,sn + 1} \right\},q' = q,i'' \in \left\{ {1,...,U} \right\},i'' \notin {{\cal E}_{L(q')}}}\\
{\left\{ {1,...,sn} \right\},q' = q,i'' \in \left\{ {U + 1,...,K} \right\}}
\end{cases}} ,
\end{equation}
\begin{equation}
\tilde {\cal S}_{{j'i'r_q}}^{\beta,{\cal{A}}}=\{1,...,\upsilon n\}.
\end{equation}

By the same argument given before, subspaces $\bar {\cal {A}}_{r_q}$, $\tilde {\cal A}_{r_q}$ and $\tilde {\cal C}_{i,r_q},i\in {\cal E}_{L(q)}$   are full rank  and linearly independent almost surely.
We can  see that:
\begin{equation}
d(\tilde {\cal{C}}_{i,r_q})={n^{{K^2} - K}}{(sn)^{QK - U^2}}{(\upsilon n)^{U^2}},\forall i \in {\cal E}_{L(q)},
\end{equation}
\begin{equation}
D_N(\tilde {\cal{C}}_{i,r_q})=\chi,
\end{equation}
so, the normalized dimension of the total  subspaces which could be demultiplexed at the NC-IR $q$-th antenna is:
\begin{equation}
\sum\limits_{i \in {{\cal E}_{L(q)}}} {D_ N(\tilde {\cal{C}}_{i,r_q})}= \left| {{{\cal E}_{L(q)}}} \right|\chi  .
\end{equation}

For $\bar {\cal{A}}_{r_q}$, as the same as proof of Theorem \ref{theorem1}, we have:
\begin{equation}
D_N(\bar {\cal{A}}_{r_q})=\max\{\Gamma,\zeta\},
\end{equation}
For $\tilde {\cal{A}}_{r_q}$, we have:
\begin{equation}
d(\tilde {\cal{A}}_{r_q})={n^{{K^2} - K}}{(sn)^{QK - U - (U - \left| {{B_q}} \right|)}}{(sn + 1)^{U - \left| {{B_q}} \right|}}{(\upsilon n)^U},
\end{equation}
\begin{equation}
D_N(\tilde {\cal{A}}_{r_q})=\chi.
\end{equation}
Thus, we can see that interference alignment equations (\ref{IA7})-(\ref{IA10}) are satisfied. 

The same as the proof of scheme 1 in Theorem \ref{theorem1}, we derive the dimension of the  whole received signal space at each receiver.
Therefore, if we define $d_{t,j}$ as the total dimension at the $j$-th receiver and $d_{t,r_q}$ as the total dimension at the $q$-th receive antenna of the NC-IR, then, we can see (\ref{d_{t,j}})-(\ref{d_{t,r_q}}) will be obtained, if we replace $W$ and ${\cal B}_q$ with $U$ and ${\cal E}_{L(q)}$, respectively.
Therefore, considering $D_{N,t,j}$ as the total normalized asymptotic dimension at the $j$-th receiver and $D_{N,t,r_q}$ as the total normalized asymptotic dimension at the $q$-th antenna of the NC-IR, we have:
\begin{equation}
D_{N,t,j}=\Gamma +\chi +\max\{\Gamma,\zeta\},\forall j \in \{1,...,U\},
\end{equation}
\begin{equation}
D_{N,t,j}=\zeta +\chi +\max\{\Gamma,\zeta\},\forall j \in \{U+1,...,K\},
\end{equation}
\begin{equation}
D_{N,t,r_q}=\left| {{{\cal E}_{L(q)}}} \right| \chi +\chi+\max\{\Gamma,\zeta\},\forall q \in \{1,...,Q\}.
\label{D_{N,t,r_q}NC-IR}
\end{equation}
Considering the parameter $T$ as (\ref{T_min}), we have:
\begin{equation}
\mathop {\lim }\limits_{n \to \infty }\frac {T}{n^{K^2-K+QK}} = \chi  + \max \left\{ {\Gamma ,\zeta } \right\} + \max \left\{ {\mathop {\max }\limits_{q \in \{ 1,...,Q\} } \left| {{{\cal E}_{L(q)}}} \right|\chi ,\zeta ,\Gamma } \right\}.
\label{eq-a1}
\end{equation}
Moreover, we have:
\begin{equation}
\mathop {\max }\limits_{q \in \{ 1,...,Q\} } \left| {{{\cal E}_{L(q)}}} \right| = \left\lceil {\frac{U}{p}} \right\rceil .
\label{eq-a2}
\end{equation}
Therefore, from (\ref{eq-a1}) and (\ref{eq-a2}), we conclude:
\begin{equation}
\mathop {\lim }\limits_{n \to \infty } \frac{T}{{{n^{{K^2} - K + QK}}}} = \chi  + \max \left\{ {\Gamma ,\zeta } \right\} + \max \left\{ {\left\lceil {\frac{U}{p}} \right\rceil \chi ,\zeta ,\Gamma } \right\}.
\end{equation}

Also, we let:
\begin{equation}
\Gamma = \zeta,
\label{assumption3}
\end{equation} 
\begin{equation}
\zeta  \ge \left\lceil {\frac{U}{p}} \right\rceil \chi.
\label{assumption4}
\end{equation}
By assumptions (\ref{assumption3}) and (\ref{assumption4}), we can see that the total normalized length is:
\begin{equation}
\mathop {\lim }\limits_{n \to \infty } \frac{T}{{{n^{{K^2} - K + QK}}}} = \chi +2\Gamma.
\end{equation}

\vspace{10pt}

\textbf{Step 6: DoF analysis}

Now, we characterize the total DoF. As  stated before, we have $U$ clean receivers each with normalized message dimension equal to $\Gamma+\chi$, and  $K-U$ dirty receivers each with normalized message dimension equal to $\zeta$ (note that we assumed $\zeta=\Gamma$). Therefore, the total normalized length of $T$ is equal to $\chi +2\Gamma$. Thus, the total DoF has the following form:
\begin{equation}
{\rm DoF} = \mathop {\max }\limits_{\chi  \ge 0,\Gamma  \ge \left\lceil {\frac{U}{p}} \right\rceil \chi } \frac{{U(\chi  + \Gamma ) + (K - U)\Gamma }}{{\chi  + 2\Gamma }}.
\label{optimization2}
\end{equation}
By assuming $\Gamma=\beta \chi$,  we have:
\begin{align}
{\rm DoF} &= \mathop {\max }\limits_{\beta  \ge \left\lceil {\frac{U}{p}} \right\rceil } \frac{{U(1 + \beta ) + (K - U)\beta }}{{1 + 2\beta }}\\
& = \frac{K}{2} + \mathop {\max }\limits_{\beta  \ge \left\lceil {\frac{U}{p}} \right\rceil } K\frac{{\frac{U}{K} - \frac{1}{2}}}{{1 + 2\beta }} = \frac{K}{2} + \max \left\{ {K\frac{{\frac{U}{K} - \frac{1}{2}}}{{1 + 2\left\lceil {\frac{U}{p}} \right\rceil }},0} \right\},
\label{main result3}
\end{align}
where (\ref{main result3}) follows from the fact that if $\frac{U}{K} > \frac{1}{2}$, we set $\beta=\left\lceil {\frac{U}{p}} \right\rceil$, and if $\frac{U}{K}<\frac{1}{2}$, we tend $\beta$ to $\infty$. This completes the proof.

\end{appendices}

\end{singlespace}
\end{document}